\pdfoutput=1
\documentclass[pra,twocolumn,showpacs,floatfix]{revtex4}
\usepackage{graphicx}
\usepackage{color}
\usepackage{amsmath}
\DeclareUnicodeCharacter{2009}{\,}
  
\begin{document}

\title{Rotating quantum droplets confined in 
a harmonic potential}

\author{S. Nikolaou$^1$, G. M. Kavoulakis$^{1,2}$, and M. \"{O}gren$^{2,3}$}
\affiliation{$^1$Hellenic Mediterranean University, P.O. Box 1939, GR-71004, Heraklion, Greece
\\
$^2$HMU Research Center, Institute of Emerging Technologies, GR-71004, Heraklion, 
Greece
\\
$^3$School of Science and Technology, \"{O}rebro University, 70182 \"{O}rebro, Sweden}
\date{\today}

\begin{abstract}

We investigate the rotational properties of a two-component, two-dimensional 
self-bound quantum droplet, which is confined in a harmonic potential and compare 
them with the well-known problem of a single-component atomic gas with contact 
interactions. For a fixed value of the trap frequency, choosing some representative 
values of the atom number, we determine the lowest-energy state, as the angular 
momentum increases. For a sufficiently small number of atoms, the angular momentum 
is carried via center-of-mass excitation. For larger values, when the angular 
momentum is sufficiently small, we observe vortex excitation instead. Depending 
on the actual atom number, one or more vortices enter the droplet. Beyond 
some critical value of the angular momentum, however, the droplet does not
accommodate more vortices and the additional angular momentum is carried via 
center-of-mass excitation in a ``mixed" state. Finally, the excitation 
spectrum is also briefly discussed.

\end{abstract}


\maketitle

\section{Introduction}

The rotational properties of trapped atomic Bose-Einstein condensates
is a problem which has been studied very extensively in the last decades.
Most of these studies have been performed in a harmonic potential, since
this has been by far the most common form of confining potential that 
is used in experiments. We stress that the literature on this problem 
is very extensive, so we simply refer to some review articles \cite{rev1, 
rev2, rev3, rev4, rev5}.

The interatomic interactions are modeled as an effective hard-core potential. 
This potential is proportional to the so-called scattering length, which 
describes the elastic, s-wave atom-atom collisions. In the single-component 
condensates, when this effective interaction is repulsive (i.e., the scattering 
length is positive), as the angular momentum increases, vortices enter the cloud 
from its periphery and eventually a vortex lattice forms. When the angular 
momentum increases even more, the system reaches the so-called limit of ``rapid 
rotation", where the mean-field approximation fails. The cloud enters a 
highly correlated regime, and its many-body state resembles a (bosonic) 
Laughlin-like state. On the other hand, when the effective interaction is 
attractive (i.e., the scattering length is negative), the cloud is unstable 
against collapse if there is no trapping potential. Still, the system may be 
in a metastable state due to the trap. In this case, the cloud carries its 
angular momentum via center-of-mass excitation of the ground (nonrotating) 
state. 
  
More recently Petrov \cite{Petrov} predicted in the case of 
a two-component Bose-Einstein condensate the existence of ``quantum 
droplets". This is a very interesting problem and has attracted a lot of 
attention, see, e.g., the review articles \cite{rrev1, rrev2}, and 
Refs.\,\cite{PA, th0, th1, th2, th3, th4, th5, th6, th7, th8, th9, th10, 
th11, th12, th13, th14, th15, th16, EK, add1, th166, add2, add3, add4, add5, add6}. Interestingly enough, such 
droplets have been observed experimentally not only in two-component
Bose-Einstein condensed gases \cite{qd7, qd8, qd8a, gd8b, qd8c} but also in 
single-component gases with strong dipolar interactions \cite{qd1, qd2, qd3, 
qd4, qd5, qd6}. 

The basic idea in the case where droplets are formed from binary mixtures 
is that, due to the fact that we have a two-component system, by tuning the 
strength of the effective interaction between the same and different components, 
the mean-field interaction energy may become as small as we wish.
In this case the next-to-leading-order correction of the energy (i.e., the 
so-called ``Lee-Huang-Yang" term) \cite{LHY}, becomes comparable with the 
usual mean-field term and the two terms may balance each other, giving rise 
to self-bound droplets, even in the absence of any trapping potential. 

Self-bound droplets belong to the class of systems which are superfluid. 
It is thus natural to examine their rotational properties. Compared with 
the problem of single-component atomic Bose-Einstein condensates, there 
are two main differences, which introduce novel effects in their superfluid
properties. First of all, as we saw earlier, while quantum droplets are 
self-bound and do not require any trapping potential, in the case of 
single-component atomic condensates, the presence of a confining potential 
is absolutely necessary. Second, in quantum droplets, the sign of the 
nonlinear term depends on the density, being attractive for sufficiently 
low densities and repulsive, for higher densities. On the other hand, in 
single-component condensates the interaction is modeled as a hard-core 
potential and it is either (purely) repulsive, or (purely) attractive.

As we explain below, the question of how a quantum droplet carries angular 
momentum is essentially trivial when there is no external confining potential. 
On the other hand, it becomes novel and interesting when the droplet is confined 
in a trapping potential \cite{th10, EK}. This is precisely the problem that we 
investigate below. More specifically, we consider a harmonically trapped 
two-dimensional ``symmetric" droplet. This consists of two components, however, 
due to the symmetry between them, the problem reduces to a single order parameter 
which is common to both of them. We minimize the energy under a fixed expectation 
value of the total angular momentum $L \hbar$, and a fixed value of the total atom 
number $N$ of the two components of the droplet.

According to the results of our study, the combination of a (harmonic) trapping 
potential with the more ``complex" nonlinear term introduces a very serious 
difference in the rotational response of a droplet, as compared with the case 
of contact interactions. For a sufficiently small $N$ the droplet executes 
center-of-mass rotation. For larger $N$ and small $L$ the droplet develops 
surface waves and eventually a single vortex enters the droplet. With increasing 
$L$, depending on the value of $N$ more vortices may enter the cloud, up to some 
critical value of $L$. Beyond this value, it is no longer energetically favorable 
for the droplet to accommodate more vortices. The additional angular momentum is 
then carried via center-of-mass excitation, in a ``mixed" state.

In Sec.\,II we present the model that we use. Then, in Sec.\,III we present and 
analyze our results for some representative values of $N$ and various values of
$L$. In Sec.\,IV we present the general picture that results from our analysis. 
In Sec.\,V we present some results from the excitation spectrum that we have 
found. In Sec.\,VI we investigate the experimental relevance of our results.
Finally, in Sec.\,VII we summarize the main results of our study and compare 
the present problem with the ``traditional" one, i.e., that of a single-component 
with an (attractive, or repulsive) effective contact interaction. 

\section{Model}

In what follows below we work with dimensionless units. In Sec.\,VI we restore the
units in order to make contact with experimentally relevant parameters. Assuming 
that there is a very tight confining potential along the axis of rotation, we consider 
motion of the atoms in the perpendicular plane, i.e., two-dimensional motion. We also 
assume that the quantum droplet is confined in a two-dimensional harmonic potential
\begin{equation}
  V(\rho) = \frac 1 2 \omega^2 \rho^2,
\end{equation} 
where $\omega$ is the frequency of the harmonic potential and $\rho$ is the radial 
coordinate in cylindrical-polar coordinates. 

As mentioned also above, we consider the ``symmetric" case, where the scattering 
lengths for the elastic atom-atom collisions between the same species are assumed 
to be equal for the two components. Also, both the masses of the two species, as well 
as the densities of the two components are equal. In this case the system is described 
by a single order parameter $\Psi(\rho, \theta)$, where $\theta$ is the angle in 
cylindrical-polar coordinates. Working with fixed $L$ and $N$, we minimize the following 
extended energy functional, \cite{GO} which, in dimensionless units, takes the form 
\cite{PA, th166}  
\begin{eqnarray}
  {\cal E}(\Psi, \Psi^*) = 
  \nonumber \\
  = \int \left( \frac {1} {2} |\nabla \Psi|^2 
  + \frac 1 2 \omega^2 \rho^2 |\Psi|^2
  + \frac 1 2 |\Psi|^4 \ln \frac {|\Psi|^2} {\sqrt{e}} \right) \, d^2 \rho 
 \nonumber \\
  - \mu \int \Psi^* \Psi \, d^2 \rho - \Omega \int \Psi^* {\hat L} \Psi \, d^2 \rho.
\label{funncc}
\end{eqnarray}
In the above equation $\Psi$ is normalized to the number of atoms, $\int |\Psi|^2 
\, d^2 \rho = N$. Also, ${\hat L}$ is the operator of the angular momentum, while $\mu$ 
and $\Omega$ are Lagrange multipliers, corresponding to the conservation of the atom 
number and of the angular momentum, respectively. 

The corresponding nonlinear equation that $\Psi(\rho, \theta)$ satisfies is
\begin{equation}
 \left( - \frac 1 2 \nabla^2 + \frac 1 2 \omega^2 \rho^2 
 + |\Psi|^2 \ln |\Psi|^2 - \Omega {\hat L} \right) \Psi = \mu \Psi. 
 \label{nlin}  
\end{equation}

\section{Rotational behavior of the droplet for various values of the atom number}

\subsection{Ground state of the droplet in the absence and in the presence
of a harmonic potential}

To understand the rotational properties of a quantum droplet in the 
presence of a harmonic confining potential, first of all, let us recall that
in the absence of any trapping potential the droplet carries its angular momentum 
via center-of-mass excitation of the ground (nonrotating) state, since this is 
a self-bound state \cite{EK}. 

For the discussion that follows it is also useful to recall that in the absence 
of a harmonic potential and in the Thomas-Fermi limit, we have the so-called 
``flat-top" droplet. The energy per particle of the droplet is, in this case,
\begin{eqnarray}
  \frac E N = \frac N {2 \pi \rho_0^2} \ln \frac {N} {\sqrt{e} \pi \rho_0^2}
  = \frac {\bar {n}} {2} \ln \frac {\bar{n}} {\sqrt{e}},
\label{eee}
\end{eqnarray}
where we have introduced the ``mean" (two-dimensional) density ${\bar n} 
= N/(\pi \rho_0^2)$. The value of the mean density of the droplet that minimizes 
the energy (which is also equal to the density of the ``flat-top" droplet, 
assumed to be constant) is ${\bar n} = N/(\pi \rho_0^2) = 1/\sqrt{e} \approx 
0.607$, while the corresponding minimum energy per particle is equal to 
$-1/(2 \sqrt{e}) \approx -0.303$. 

In the presence of a harmonic potential, in addition to the size of the 
droplet $\rho_0$ that we introduced above, we also have the oscillator length 
$a_{\rm osc} = 1/\sqrt{\omega}$. If the size of the droplet is much smaller 
than the oscillator length, $\rho_0 \ll a_{\rm osc}$ (i.e., for sufficiently
small values of $N$, or $\omega$), we still have center-of-mass excitation. 
We stress at this point that a unique feature of the harmonic potential is 
that the center-of-mass coordinate decouples from the relative coordinates, 
which is crucial for the results presented below \cite{WGS, BM, PP}. In the 
opposite limit, $\rho_0 \gg a_{\rm osc}$ (i.e., for sufficiently large values 
of $N$, or $\omega$), the rotational properties of the droplet are determined 
by the harmonic potential, where singly quantized vortices carry the angular 
momentum. 

Let us get an estimate about how $N$ and $\omega$ relate in the cross-over 
regime. From the expression $\rho_0 = [N/(\pi {\sqrt e})]^{1/2}$ that we 
mentioned above, which is valid in the Thomas-Fermi regime with no external 
potential, in order for $\rho_0$ to be equal to $a_{\rm osc}$, $N \omega 
\approx \pi \sqrt{e} \approx 5.18$. 

\begin{figure}
\includegraphics[width=\columnwidth ,angle=-0]{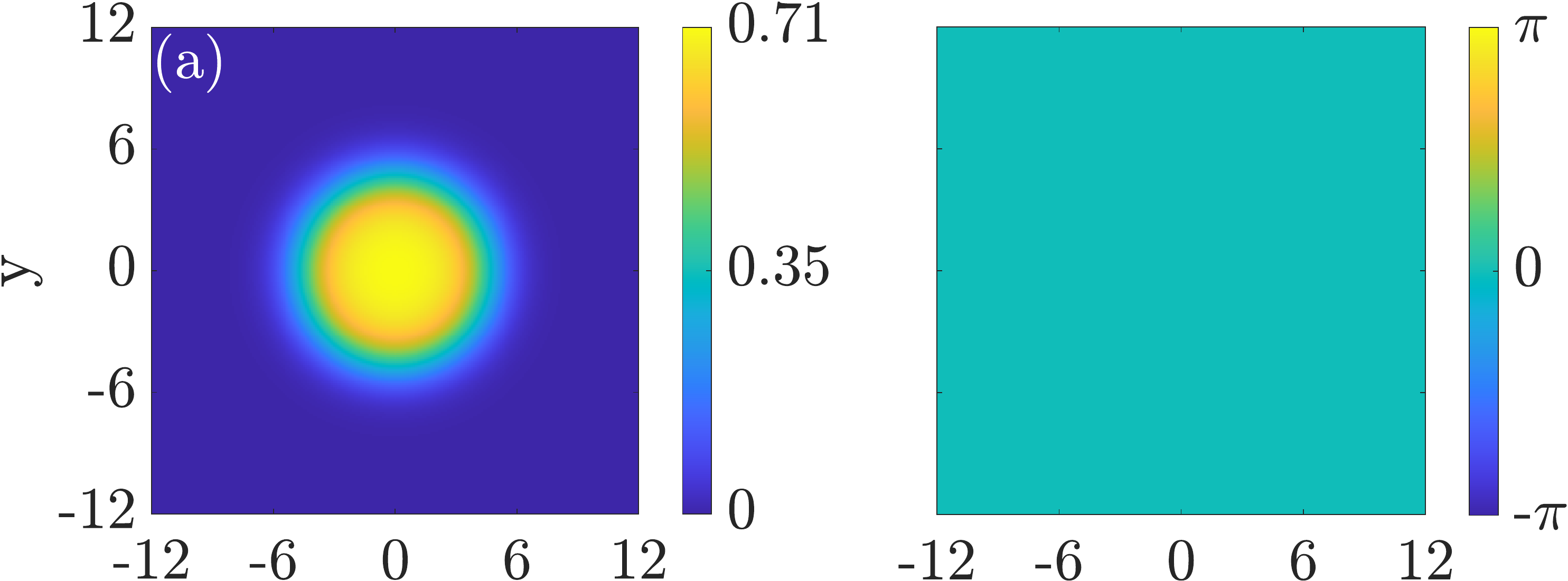}\\
\vspace{0.3\baselineskip}
\includegraphics[width=\columnwidth ,angle=-0]{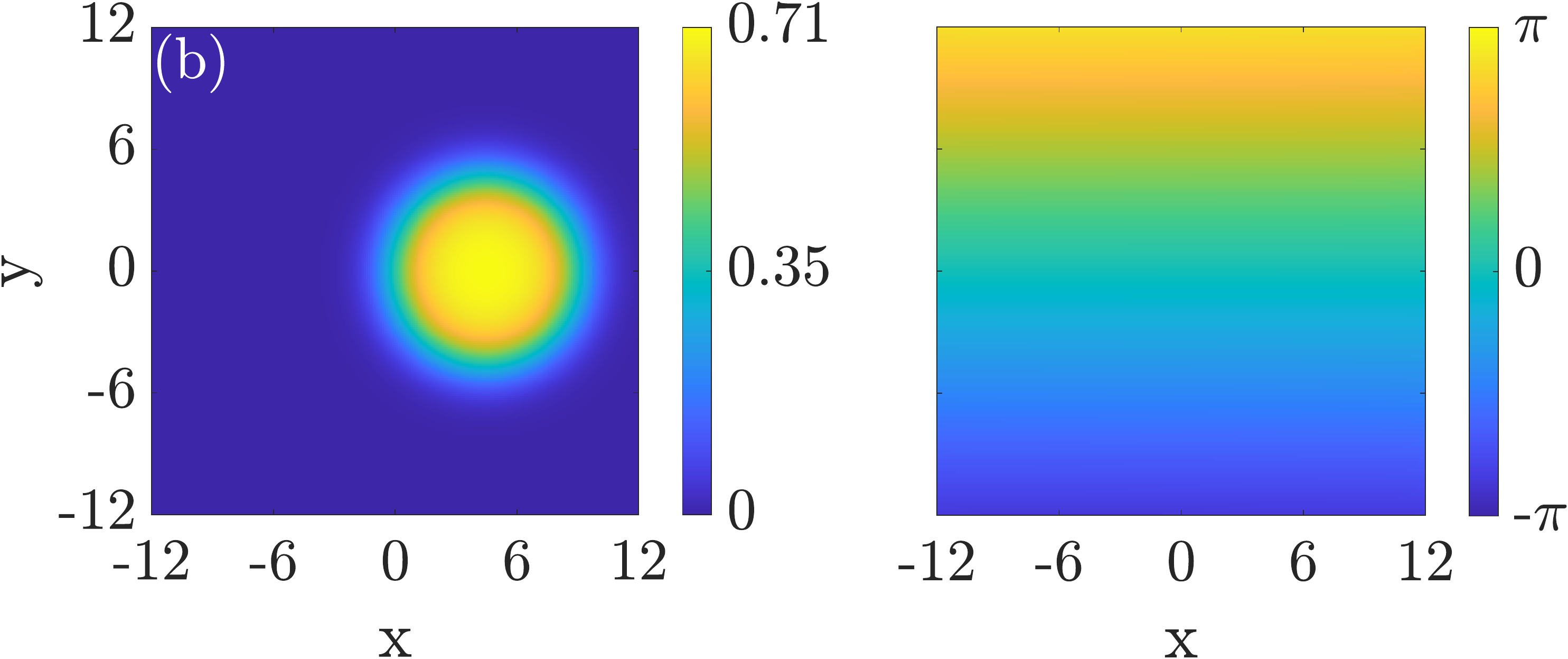}\\
\vspace{\baselineskip}
\includegraphics[width=\columnwidth ,angle=-0]{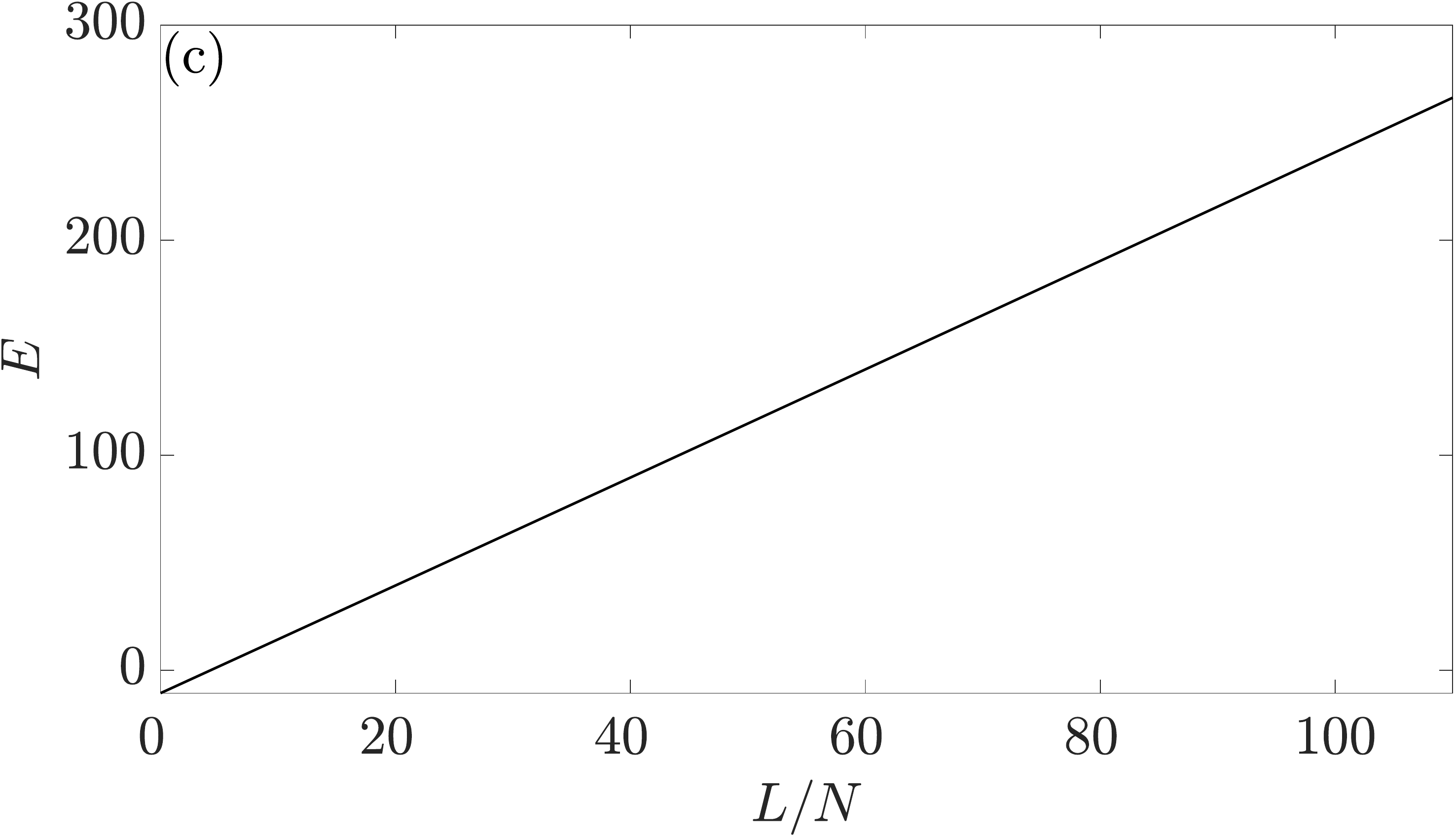}
\caption{[(a), (b)] The density (left column, in units of $\Psi_0^2$) and the phase (right column) of the droplet 
order parameter, in the lowest-energy state, for $N = 50$, $\omega = 0.05$, 
and $L/N = 0.0$ and 1.0, respectively. The unit of length is $x_0$. (c) The corresponding dispersion relation, i.e., $E = E(L/N)$. The unit of energy is $E_0$ and the unit of angular momentum is $\hbar$.}
\end{figure}

\begin{figure}[!ht]
\includegraphics[width=\columnwidth ,angle=-0]{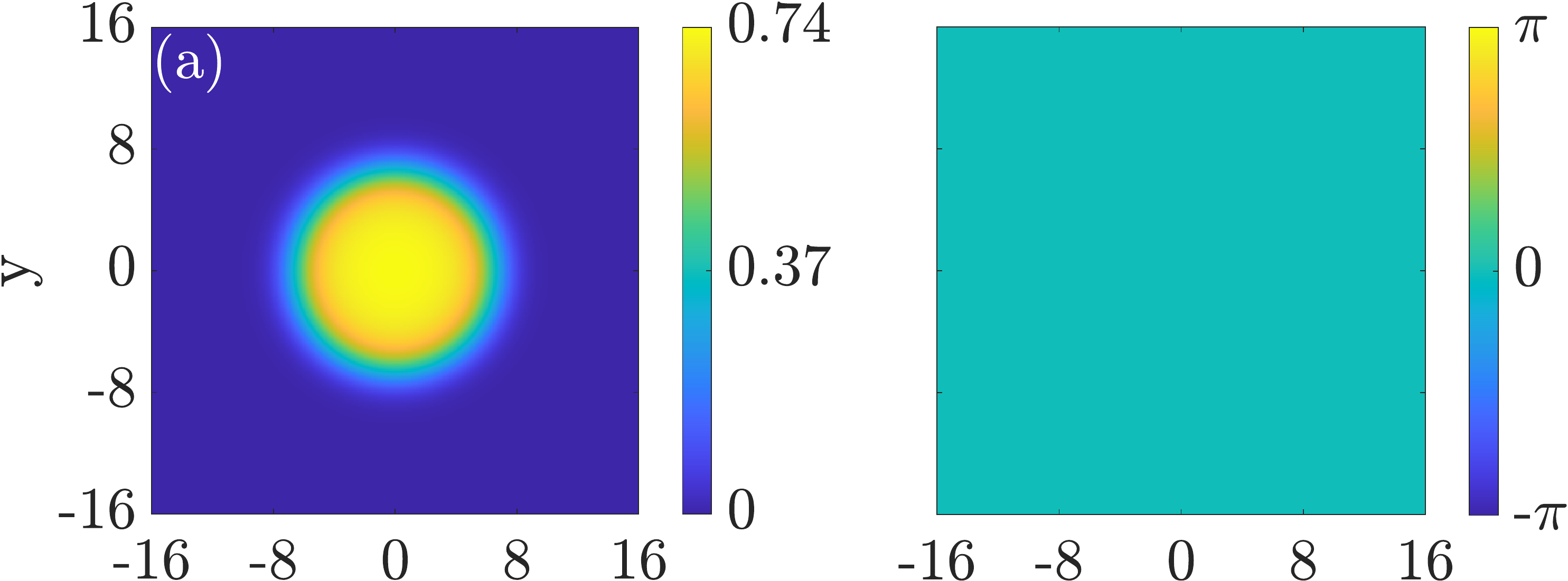}\\
\vspace{0.3\baselineskip}
\includegraphics[width=\columnwidth ,angle=-0]{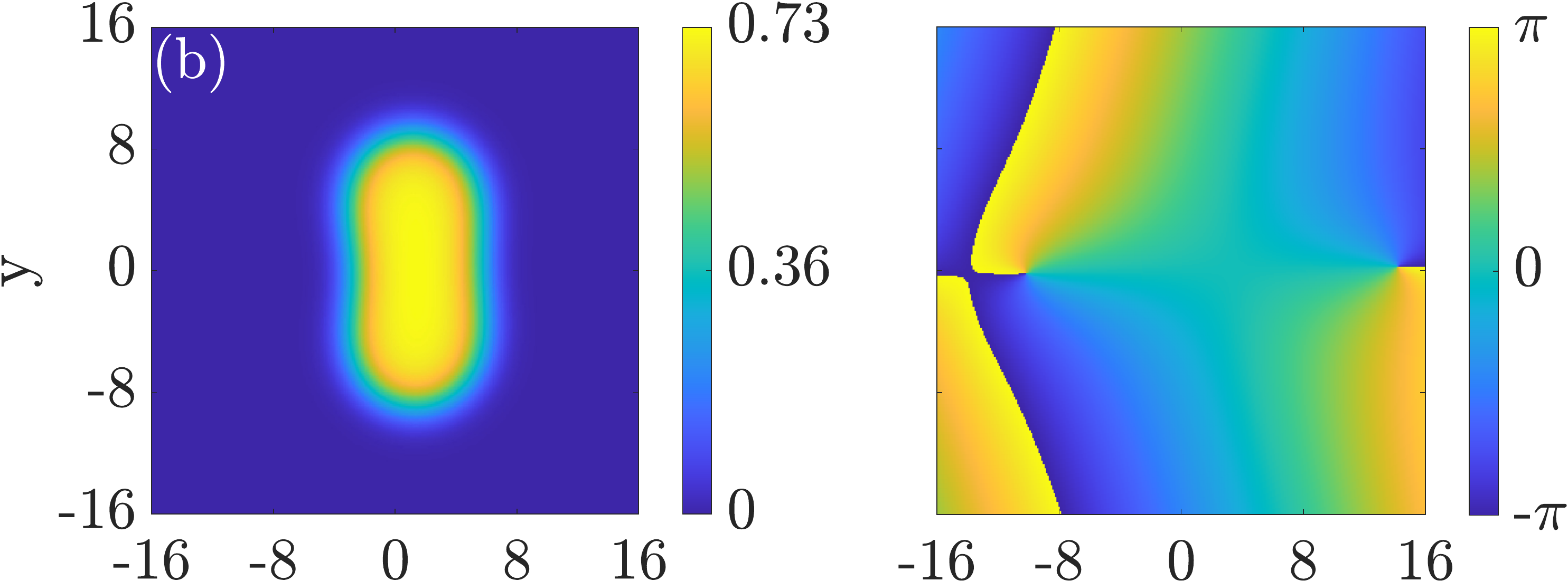}\\
\vspace{0.3\baselineskip}
\includegraphics[width=\columnwidth ,angle=-0]{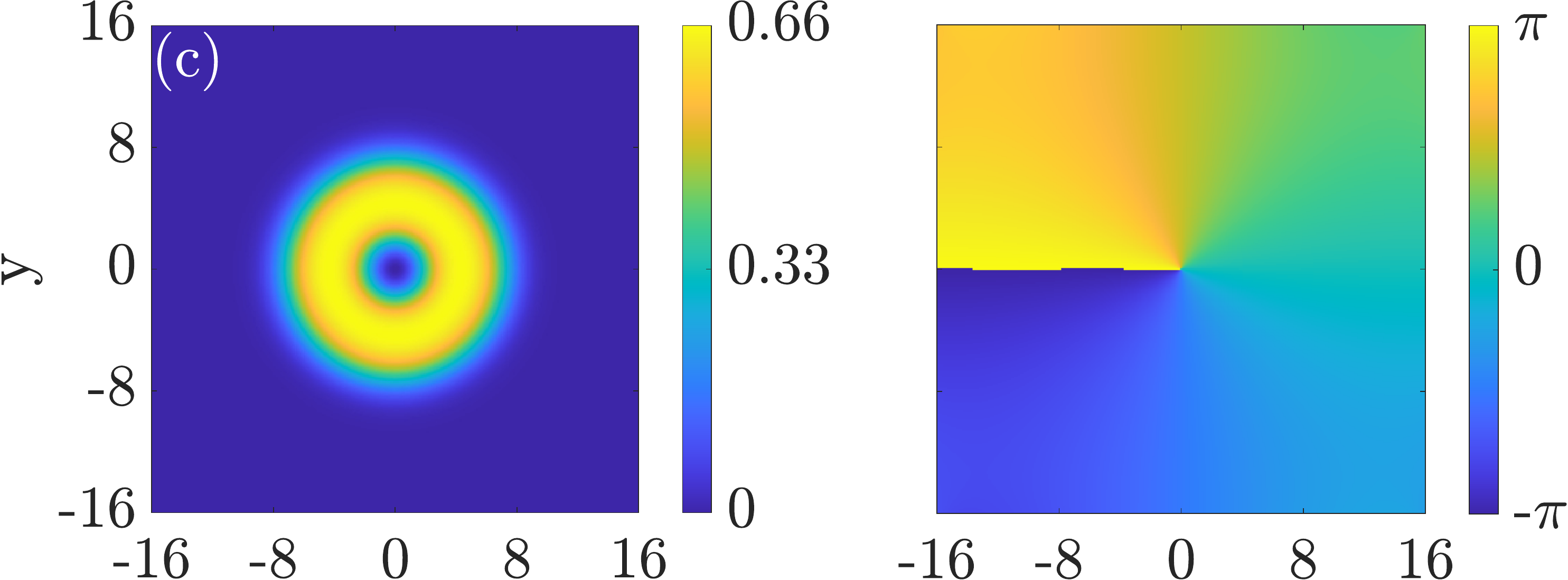}\\
\vspace{0.3\baselineskip}
\includegraphics[width=\columnwidth ,angle=-0]{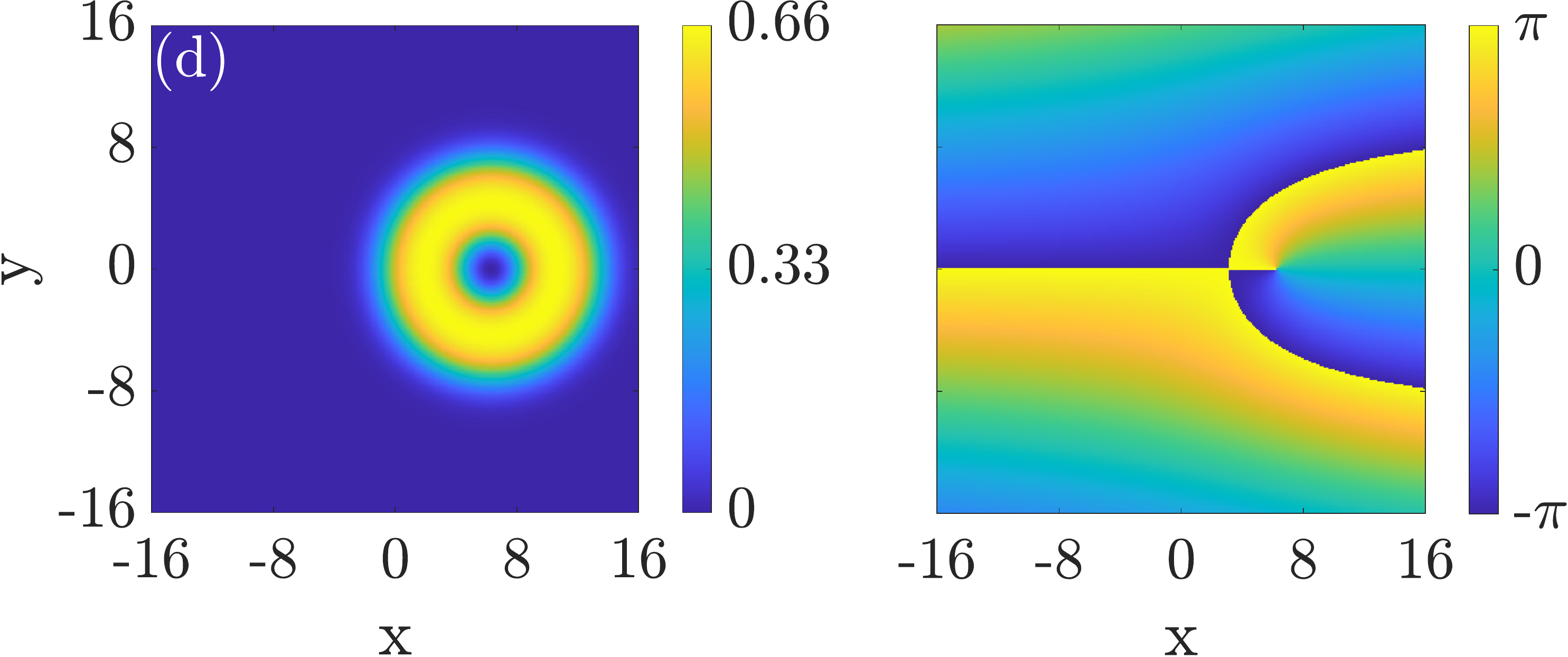}\\
\vspace{\baselineskip}
\includegraphics[width=\columnwidth ,angle=-0]{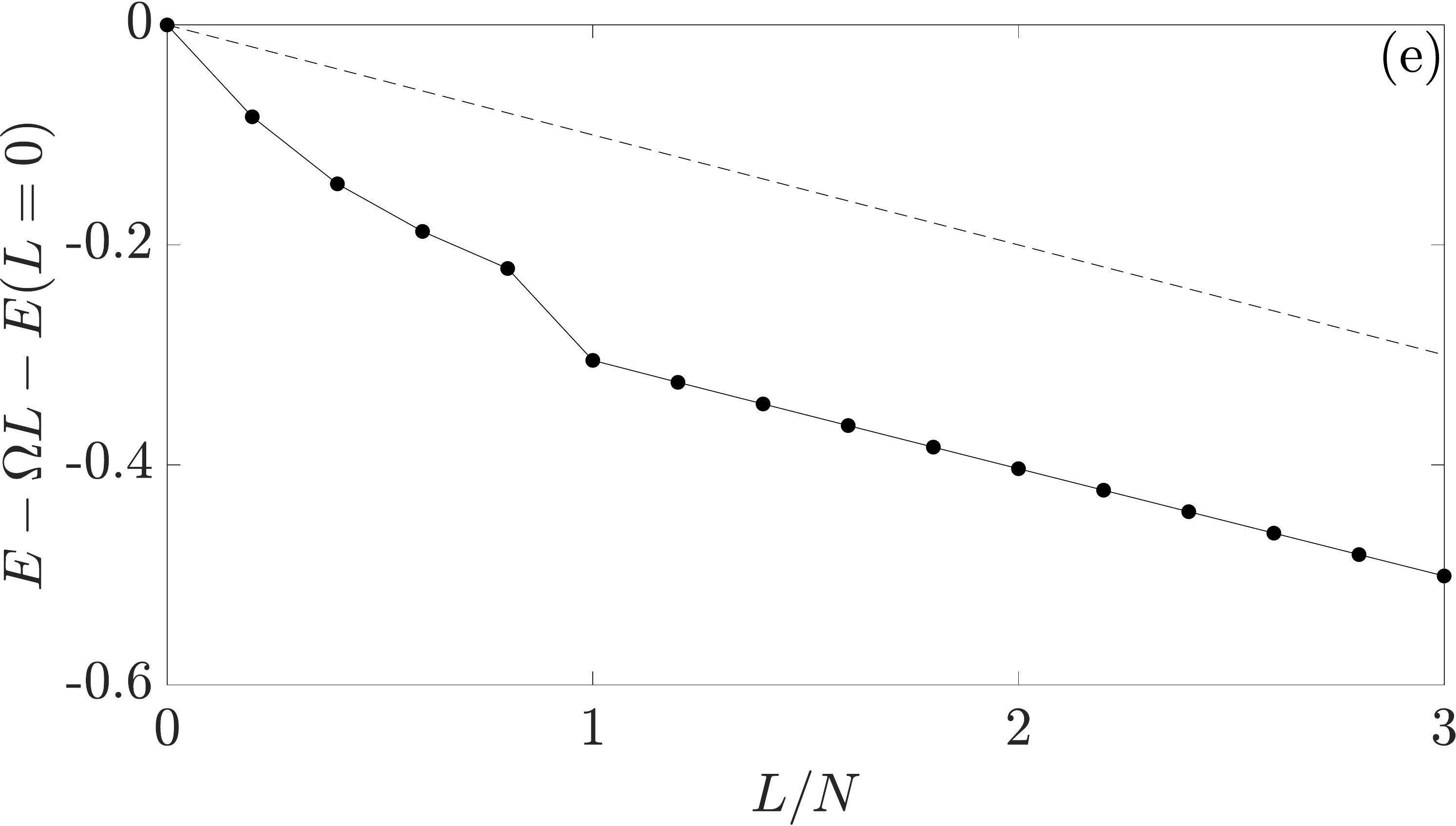}
\caption{[(a)\textendash(d)] The density (left column, in units of $\Psi_0^2$) and the phase (right column) of the droplet 
order parameter, in the lowest-energy state, for $N = 100$, $\omega = 0.05$, 
and $L/N = 0.0, 0.6, 1.0$, and 3.0, respectively. The unit of length is $x_0$. (e) Solid line, with data points: 
The corresponding dispersion relation in the rotating frame, i.e., 
$E_{\rm rot}(L/N) - E(L/N = 0)$ as function of $L/N$, with $\Omega = 0.051$. 
Dashed line: Same as above for the center-of-mass excitation of the nonrotating 
state. The unit of energy is $E_0$ and the unit of angular momentum is $\hbar$.}
\end{figure}

We minimized numerically the functional of Eq.\,(\ref{funncc}) using the 
damped second-order in fictitious time method, described in Ref.\,\cite{GO}, 
which is a method of constrained minimization. In the calculations that we 
performed, a square spatial grid was used, with $\delta x = \delta y = 0.1$. 
We checked that the choice of this grid step size gives results that are
converged with respect to the grid resolution. We stress that the actual 
size of the domain in the calculations was larger than shown in the figures, 
to avoid boundary effects. For each value of the angular momentum, 
a variety of states was used as initial conditions, to ensure that the 
calculation converged to the lowest-energy state. First, the initial 
condition for each value of angular momentum was chosen to be the converged 
solution for the previous value of the angular momentum, e.g., for $L/N=2.8$ 
the initial condition was chosen to be the converged solution for $L/N=2.6$. 
In addition to that, we repeated the calculations with different initial 
conditions, using states that represent center-of-mass excitation, surface-wave 
excitation, vortex excitation and linear combinations of these. The convergence 
of the calculation to the same solution for the majority of the chosen initial 
conditions was a strong indication that we reached the lowest-energy state for 
each value of the angular momentum.

In what follows below we present the results for four different
values of $N$, for $N = 50$ (droplets of ``small" size), $N = 100$ and $N = 200$ 
(droplets of ``intermediate" size), and $N = 270$ (droplets of ``larger" size).
These values of $N$ were chosen as representative in the sense that they give
the more general picture of this problem, which has a rather rich structure.

\subsection{Rotational properties of droplets of ``small" size}

Varying $L/N$ between 0 and 110 we show in Fig.\,1 the result of such a calculation, 
for the density and the phase of the order parameter, as well as for the energy $E(L)$,
with $\omega = 0.05$ and $N = 50$, i.e., $N \omega = 2.5$. Fitting the energy with a 
quadratic polynomial, we find that 
\begin{eqnarray}
E(L) \approx 
-10.6375 + 0.050002 \, L + 6.378 \times 10^{-8} \, L^2. 
\label{fit}
\end{eqnarray}
Both from the density [Figs.\,1(a) and 1(b)], as well as from the dispersion relation [Fig.\,1(c)], it is clear that 
we have center-of-mass excitation of the droplet for these values of $\omega$ 
and $N$. The constant term $-10.6375$ in Eq.\,(\ref{fit}) is the energy of the 
nonrotating state. Equation (\ref{eee}) gives a total energy which is $\approx 
-15.1633$. This, combined with the zero-point energy of the harmonic potential 
in two dimensions, i.e., $N \omega$, gives $-12.6633$. This number deviates from 
the numerical result $-10.6375$ and is lower due to the fact that for $N = 50$ 
the system has not yet reached the Thomas-Fermi limit and the (neglected) kinetic 
energy is not negligible. Turning to the term which is linear in $L$ in 
Eq.\,(\ref{fit}), this is due to the harmonic potential, while the term which 
is quadratic in $L$ is negligible. In other words, the more general result for 
$E(L)$ is, in this regime,
\begin{eqnarray}
  E(L) = E_{\rm COM}(L) = E(L = 0) + L \omega.
\label{eqcom}
\end{eqnarray}
We stress that Eq.\,(\ref{eqcom}) provides an upper bound for the energy, for
any value of $N$ and $L$, as we explain in more detail below.

\subsection{Rotational properties of droplets of ``intermediate" size}

For fixed $\omega$ and larger values of $N$ the size of the droplet becomes 
comparable with $a_{\rm osc}$, $\rho_0 \approx a_{\rm osc}$. In this case 
the droplet starts to get ``squeezed" due to the trapping potential. Thus, 
the trapping potential tends to increase the mean value of the density of 
the droplet, ${\bar n}$. This, in turn, increases the energy due to the 
nonlinear term, too [see Eq.\,(\ref{eee})]. In the presence of a vortex 
state ${\bar n}$ drops and therefore a vortex state may be energetically 
favorable. Indeed, as we have also seen numerically, as $N$, or as 
$\omega$, increase, we have vortex, rather than center-of-mass excitation
of the droplet. 

Such an example is shown in Fig.\,2, where $N = 100$ and $\omega = 0.05$,
i.e., $N \omega = 5$. Here we see that for small values of $L$ the axial
symmetry of the droplet is distorted [Fig.\,2(b)]. This is due to the fact that two 
vortices approach the droplet from opposite sides, with one being further 
away from the trap center than the other. Eventually, when $L = N$ the 
vortex state that is closer moves to the center of the trap and the
density of the droplet becomes axially symmetric [Fig.\,2(c)]. For even larger values 
of $L$, $L > N$, however, instead of more vortices entering the cloud, the 
extra angular momentum is carried via center-of-mass excitation of the state 
with $L = N$, i.e., the state with one vortex located at the center of 
the droplet, as shown in Fig.\,2(d). This is in sharp contrast with the case of contact interactions. 
It is a generic result and is one of the novel aspects of the present study.

The corresponding dispersion relation is also shown in Fig.\,2(e). Instead of 
plotting it in the laboratory frame, we choose to plot it in the rotating frame 
(in this plot and in all the other plots of the dispersion relation that 
follow below), because its structure is more clearly visible. More 
specifically, we plot $E_{\rm rot}(L/N) - E(L/N = 0)$, where $E_{\rm rot}(L/N) 
= E(L/N) - L \Omega$, with $\Omega = 0.051$ (i.e., we choose a slightly larger 
value of $\Omega$ than $\omega = 0.05$). When $L > N$, we see that the dispersion 
relation becomes linear, as expected, since the nonlinear term of the energy 
is unaffected by the angular momentum in this range of $L$ (simply because 
the shape of the droplet does not depend on $L$ in this range of $L$). 

\begin{figure}[!ht]
\includegraphics[width=\columnwidth ,angle=-0]{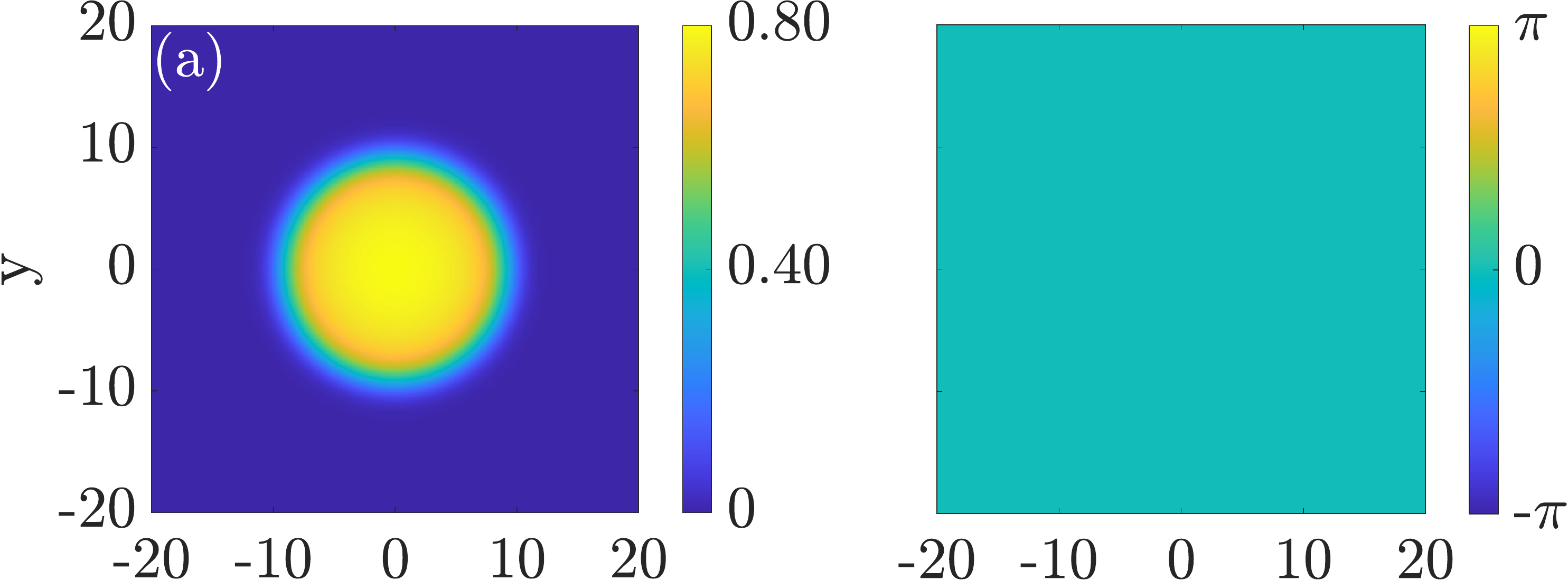}\\
\vspace{0.3\baselineskip}
\includegraphics[width=\columnwidth ,angle=-0]{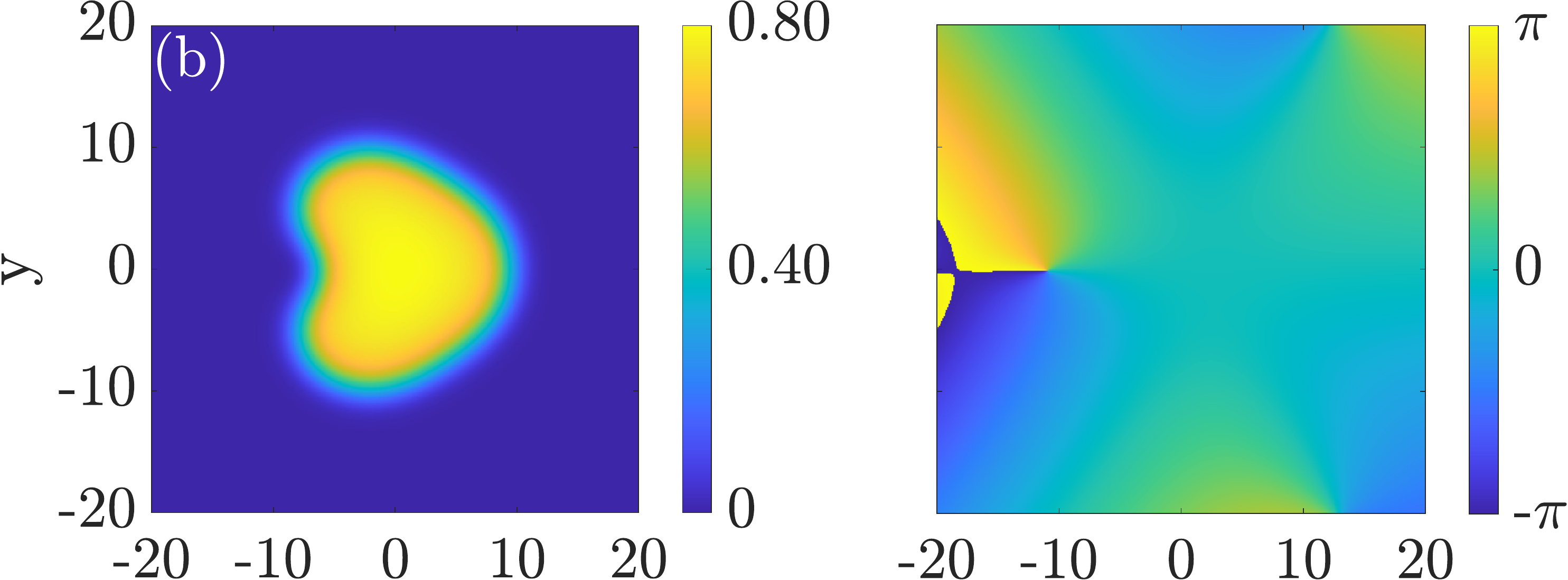}\\
\vspace{0.3\baselineskip}
\includegraphics[width=\columnwidth ,angle=-0]{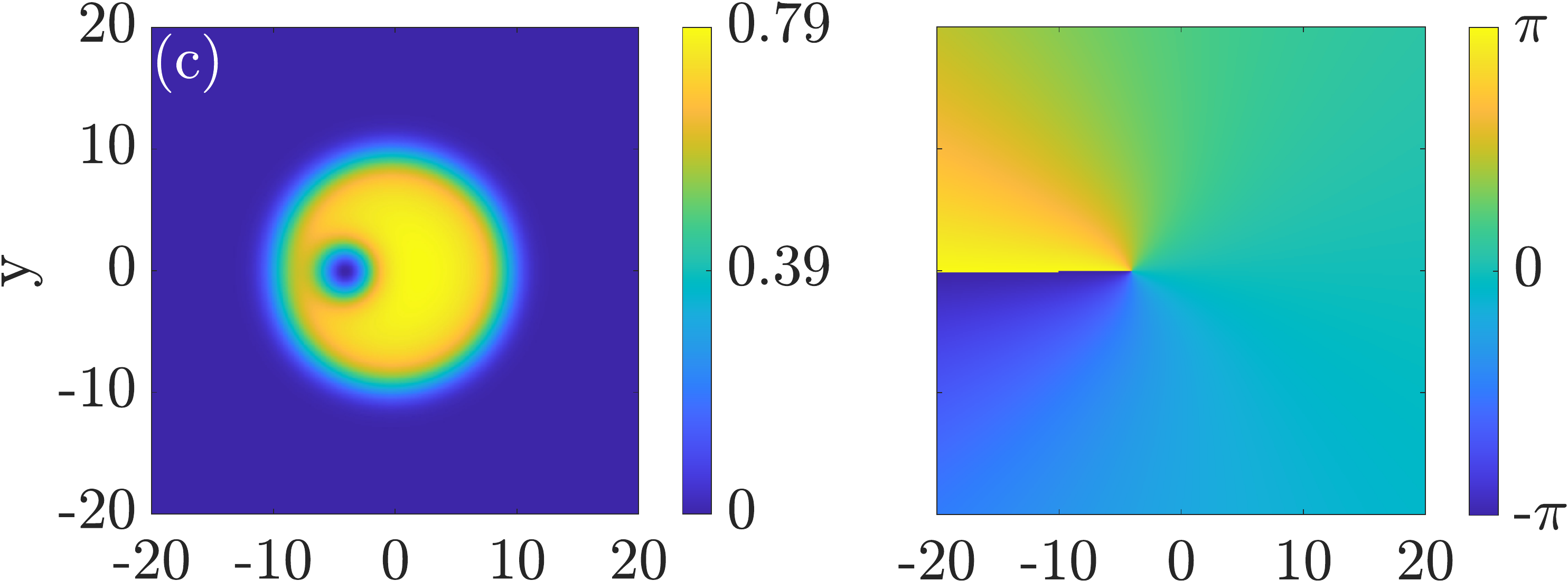}\\
\vspace{0.3\baselineskip}
\includegraphics[width=\columnwidth ,angle=-0]{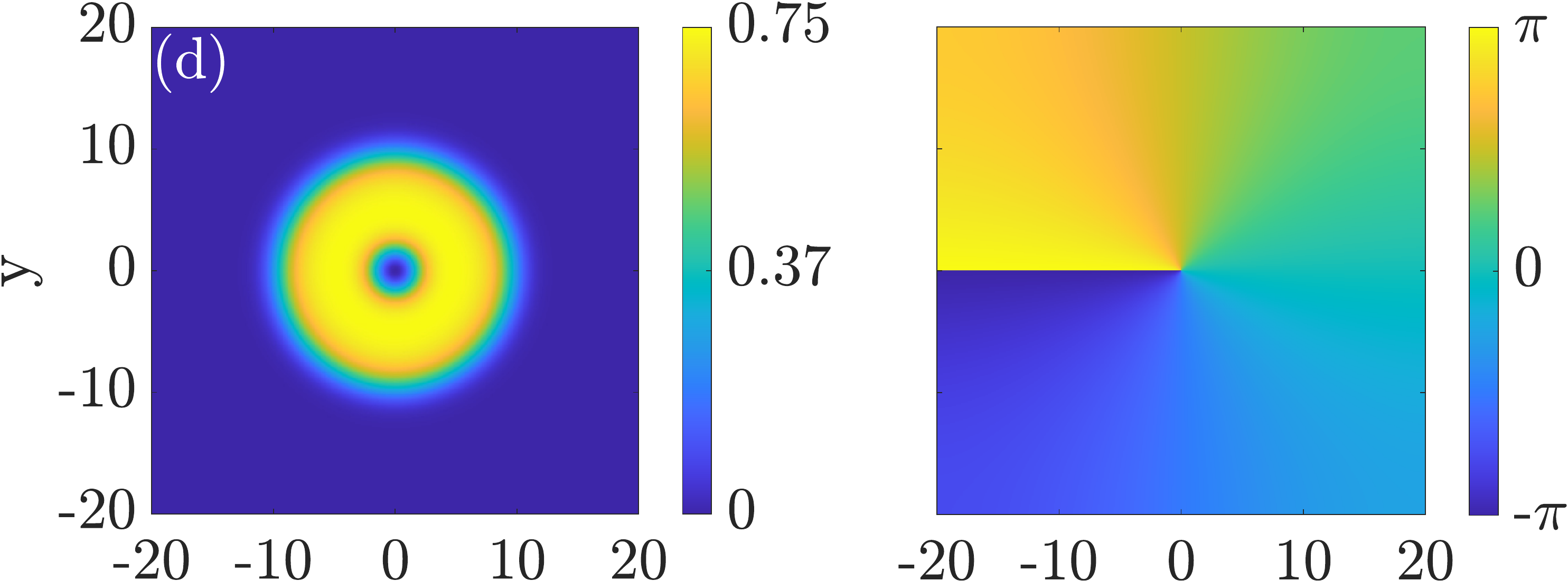}\\
\vspace{0.3\baselineskip}
\includegraphics[width=\columnwidth ,angle=-0]{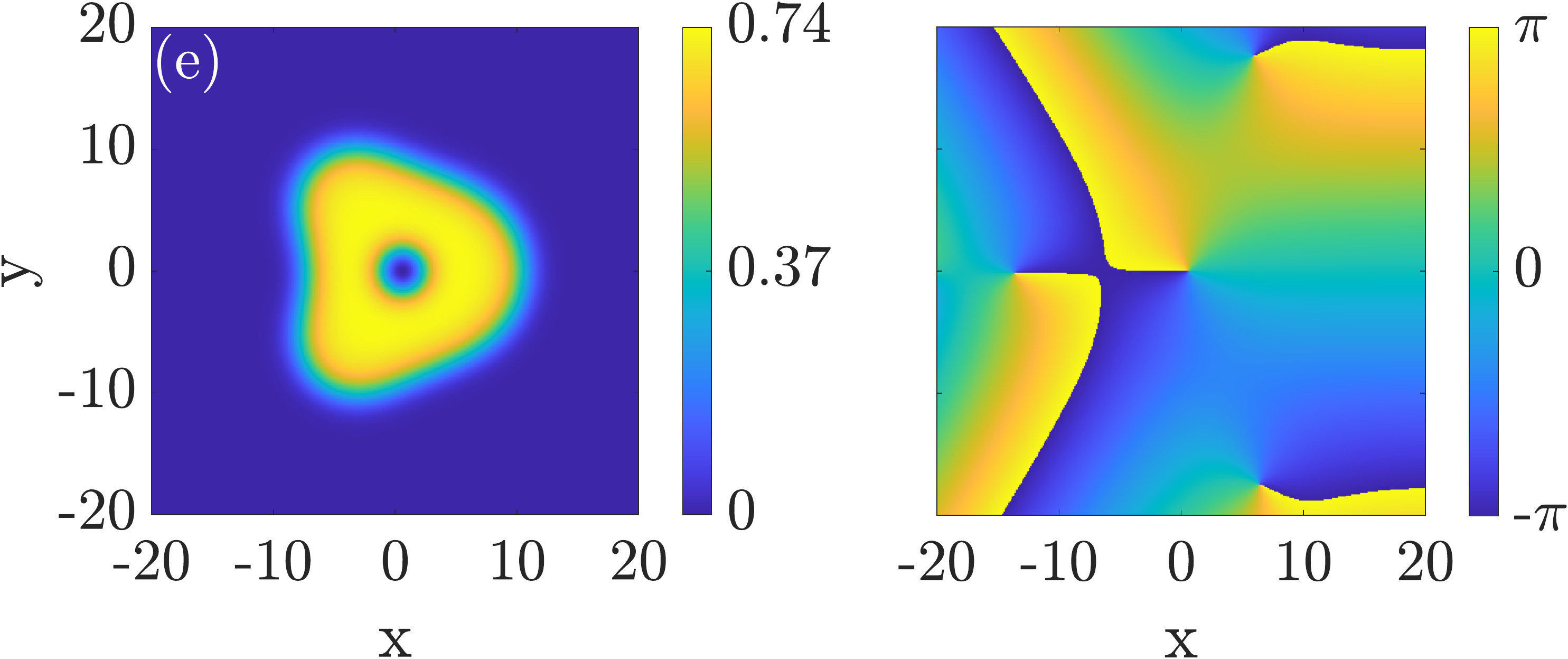}
\caption{[(a)\textendash(e)] The density (left column, in units of $\Psi_0^2$) and the phase (right column) of the droplet order parameter, in the lowest-energy state, for $N = 200$, $\omega = 0.05$, and $L/N = 0.0, 0.2, 0.8, 1.0$, and 1.2, respectively. The unit of length is $x_0$.}
\addtocounter{figure}{-1}
\end{figure}
\begin{figure}[!ht]
\includegraphics[width=\columnwidth ,angle=-0]{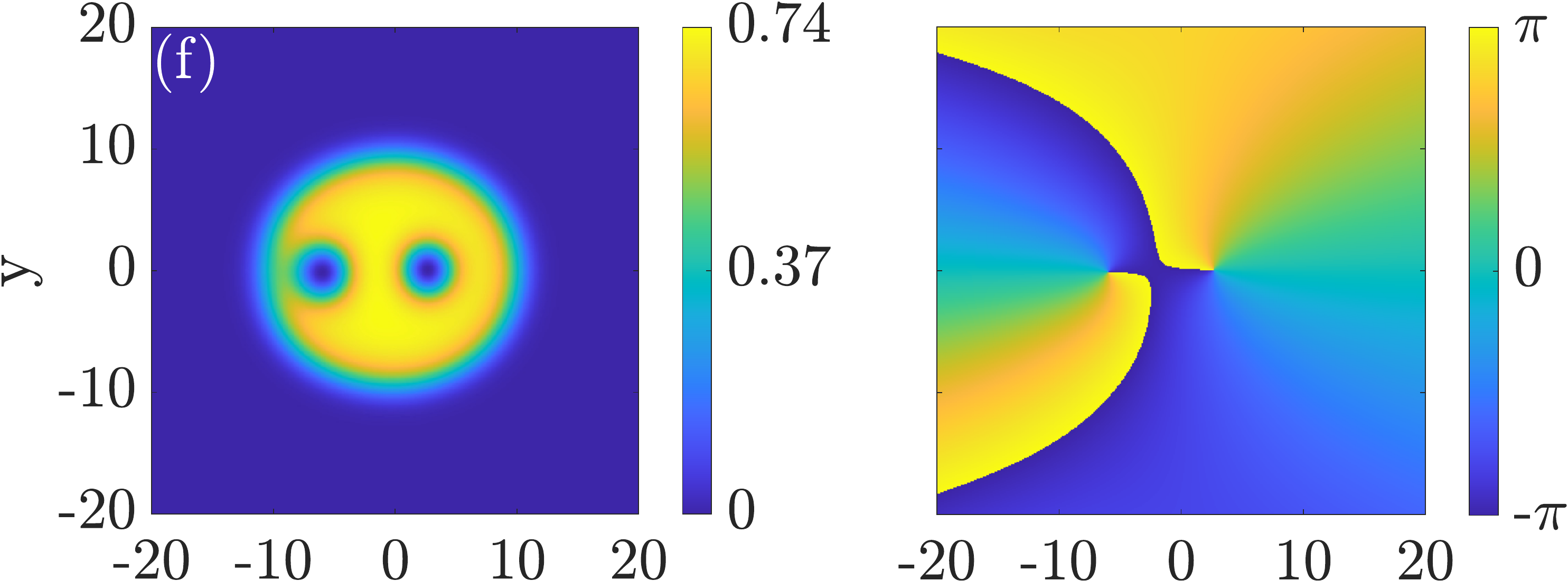}\\
\vspace{0.3\baselineskip}
\includegraphics[width=\columnwidth ,angle=-0]{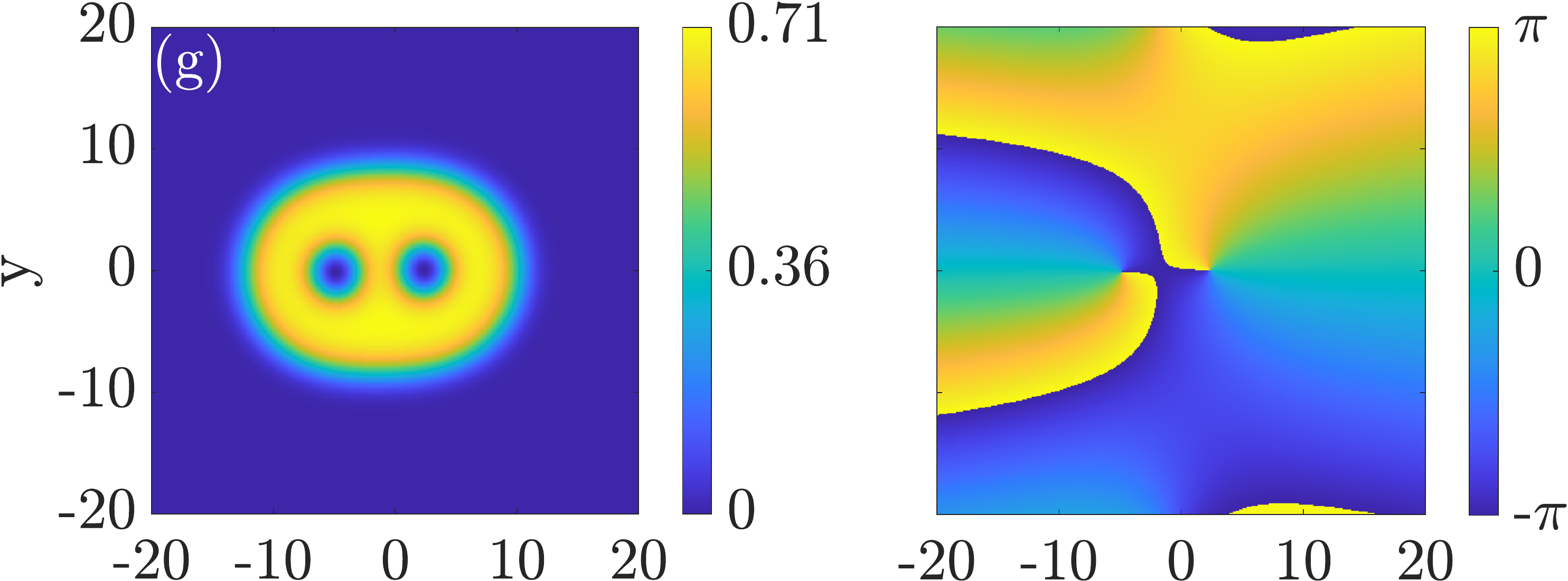}\\
\vspace{0.3\baselineskip}
\includegraphics[width=\columnwidth ,angle=-0]{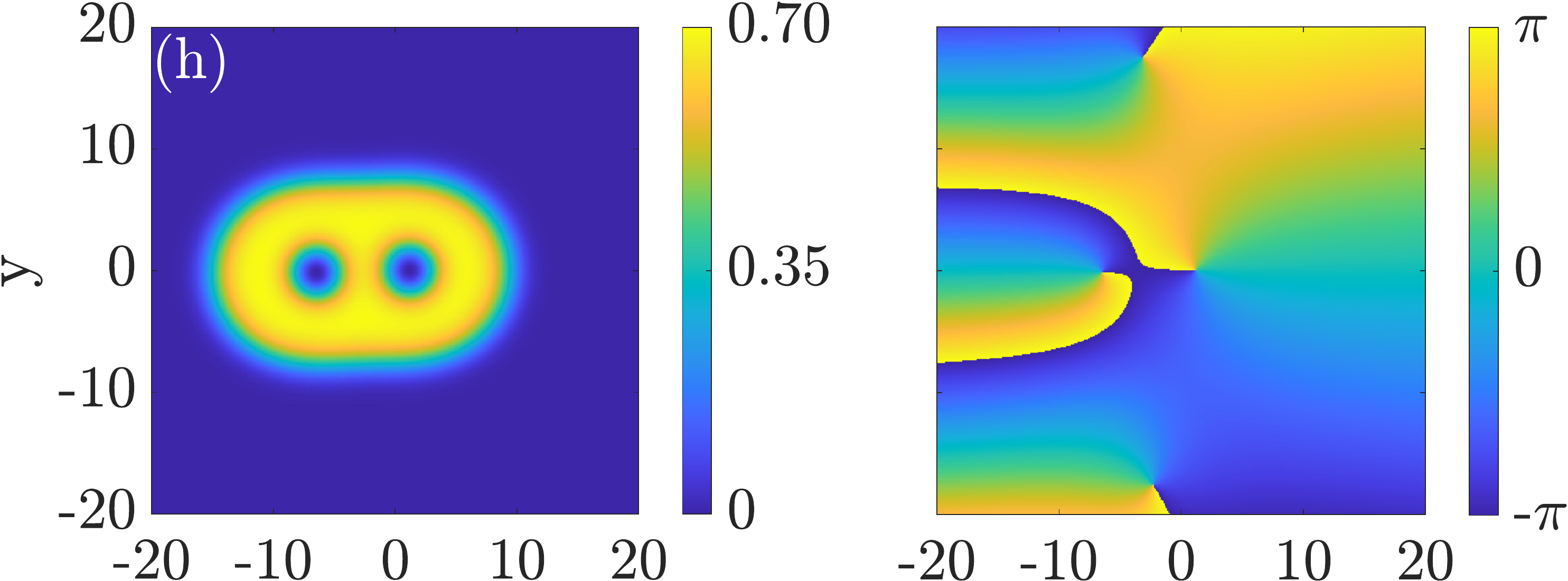}\\
\vspace{0.3\baselineskip}
\includegraphics[width=\columnwidth ,angle=-0]{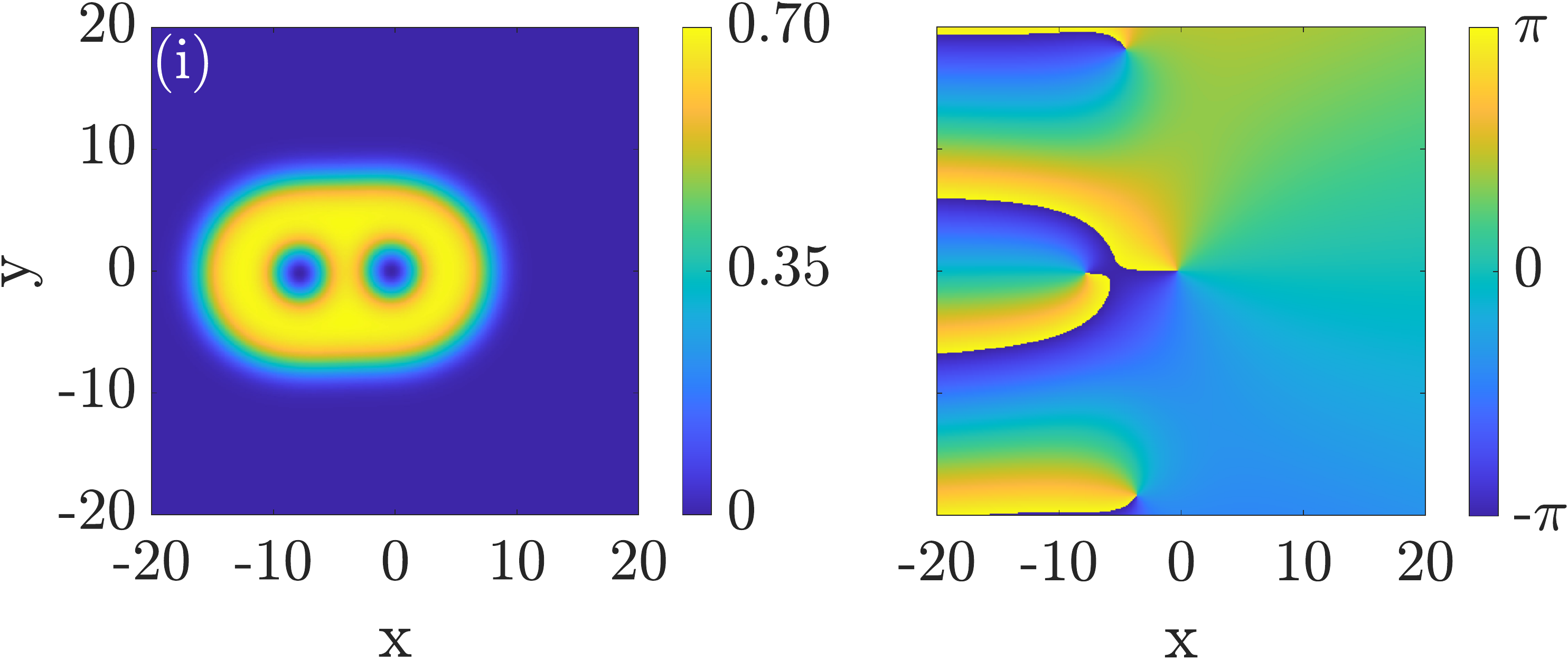}\\
\vspace{\baselineskip}
\includegraphics[width=\columnwidth ,angle=-0]{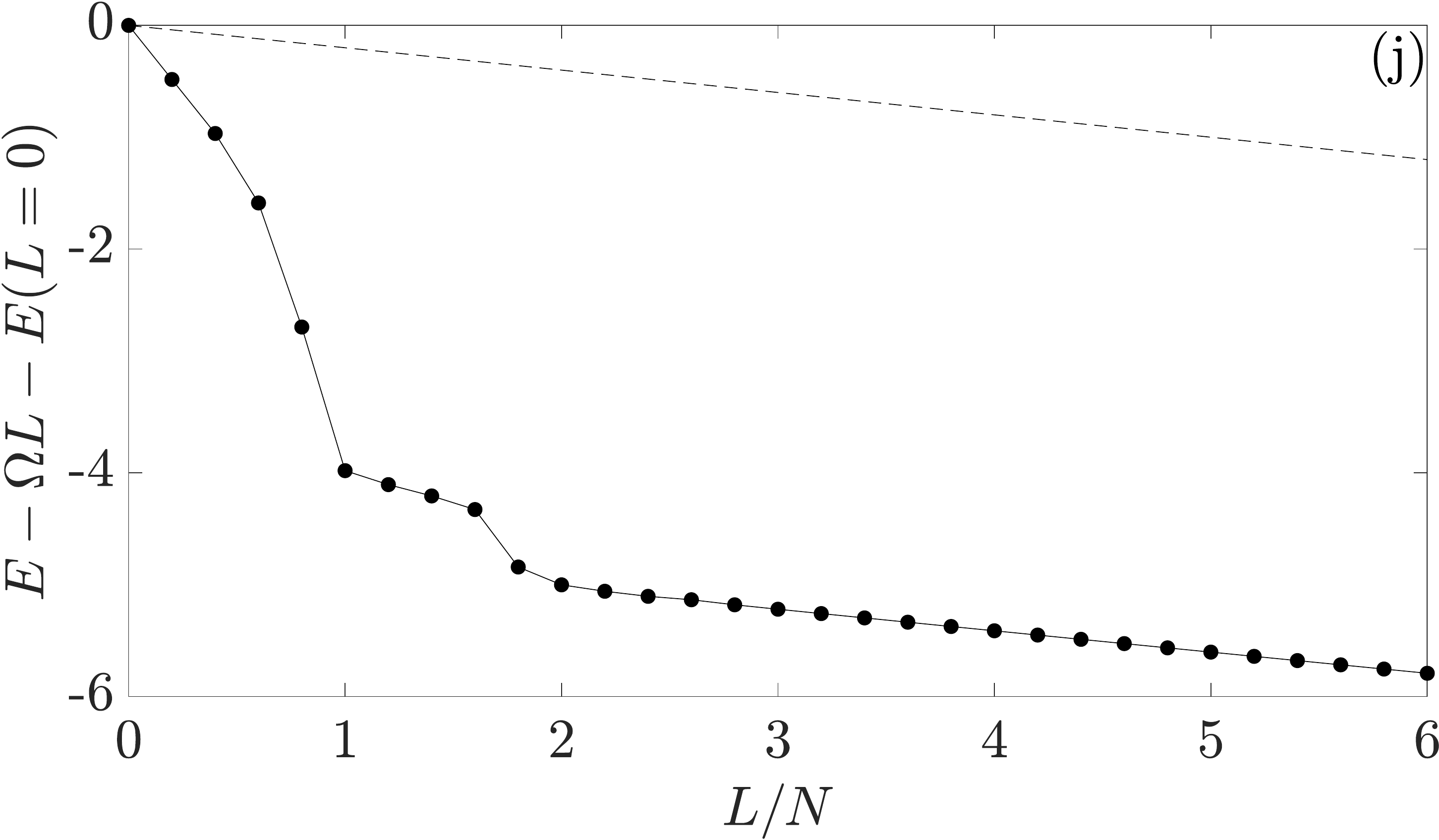}
\caption[]{(Cont.) [(f)\textendash(i)] Same as panels [(a)\textendash(e)] except for $L/N = 1.6, 2.0, 2.6$, and 3.0, respectively. (j) Solid line, with data points: the corresponding dispersion relation in the rotating frame, i.e., $E_{\rm rot}(L/N) - E(L/N = 0)$ as a function of $L/N$, with $\Omega = 0.051$. Dashed line: same as above for the center-of-mass excitation of the nonrotating state. The unit of energy is $E_0$ and the unit of angular momentum is $\hbar$.}
\end{figure}

\begin{figure}[!h]
\includegraphics[width=\columnwidth ,angle=-0]{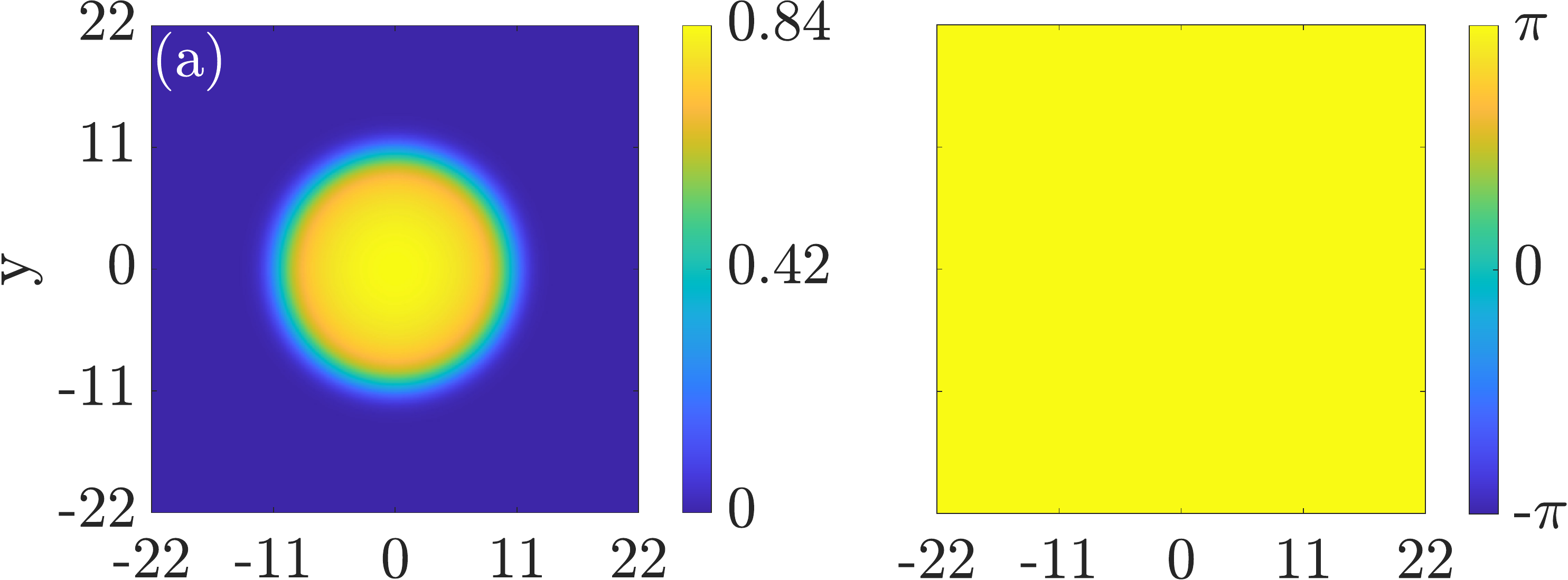}\\
\vspace{0.3\baselineskip}
\includegraphics[width=\columnwidth ,angle=-0]{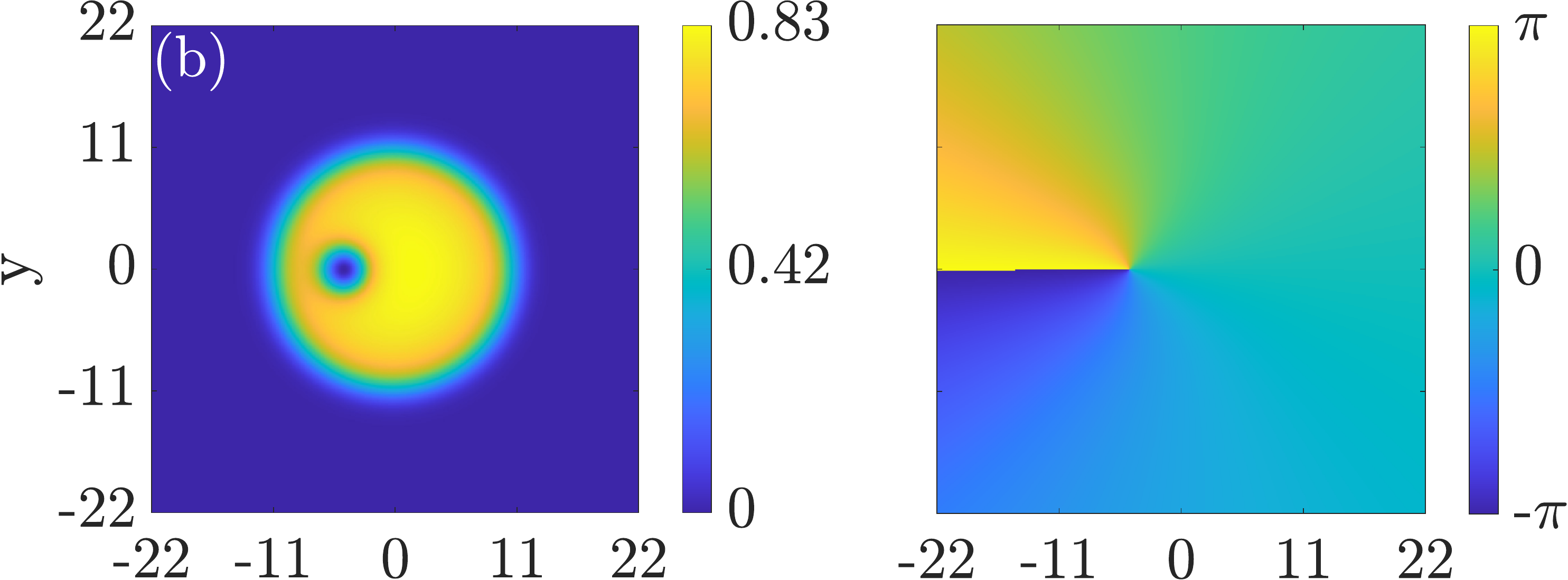}\\
\vspace{0.3\baselineskip}
\includegraphics[width=\columnwidth ,angle=-0]{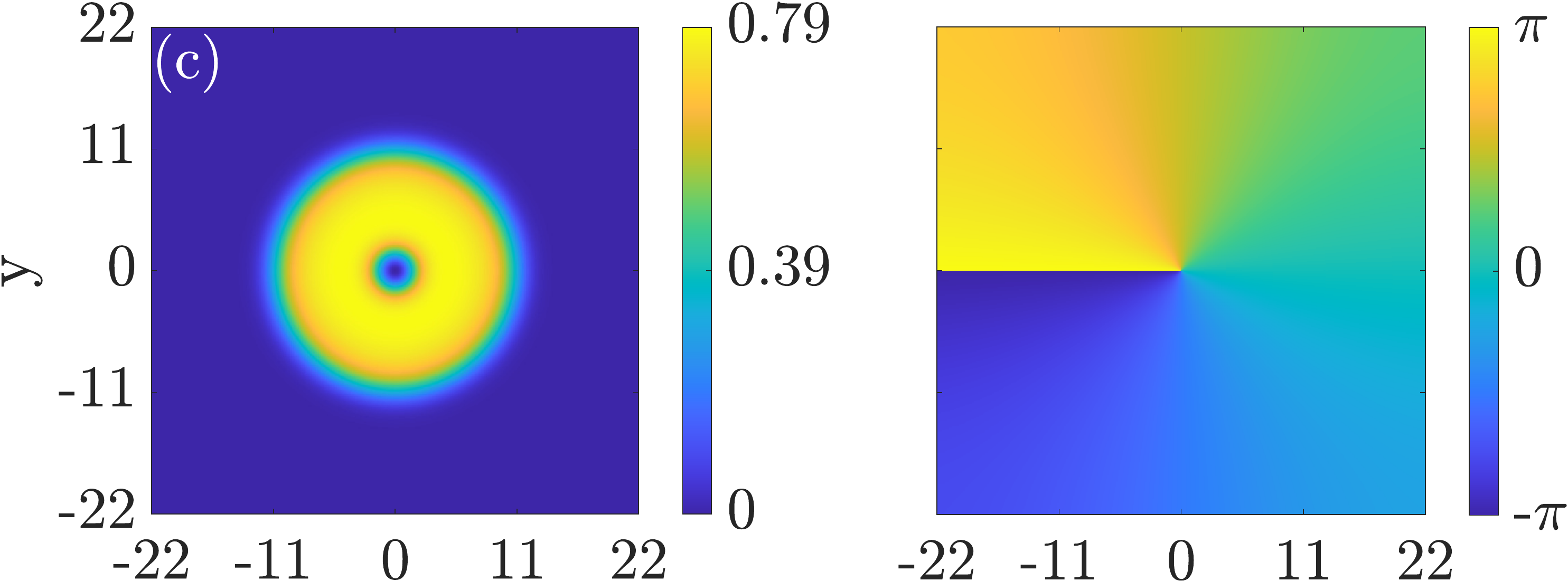}\\
\vspace{0.3\baselineskip}
\includegraphics[width=\columnwidth ,angle=-0]{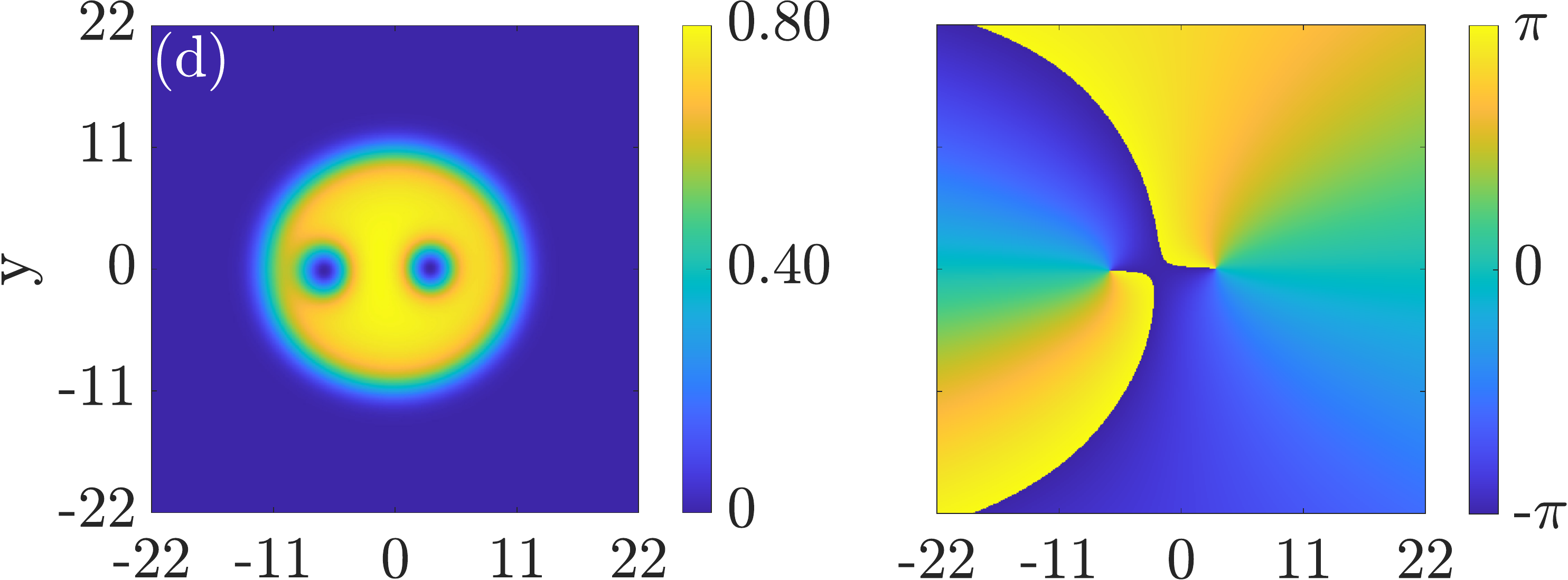}\\
\vspace{0.3\baselineskip}
\includegraphics[width=\columnwidth ,angle=-0]{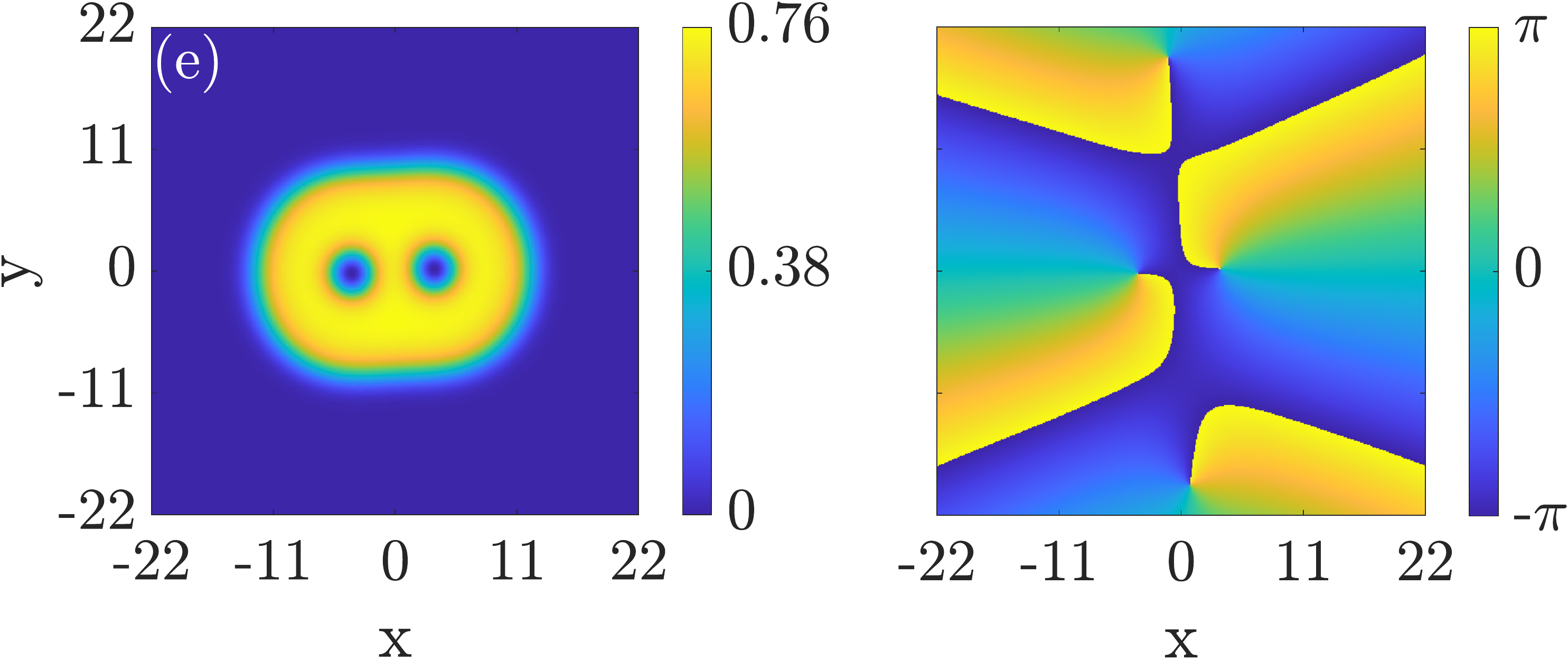}
\caption{[(a)\textendash(e)]The density (left column, in units of $\Psi_0^2$) and the phase (right column) of the droplet order parameter, in the lowest-energy state, for $N = 270$, $\omega = 0.05$, and $L/N = 0.0, 0.8, 1.0, 1.6$, and 2.0, respectively. The unit of length is $x_0$.}
\addtocounter{figure}{-1}
\end{figure}
\begin{figure}[!ht]
\includegraphics[width=\columnwidth ,angle=-0]{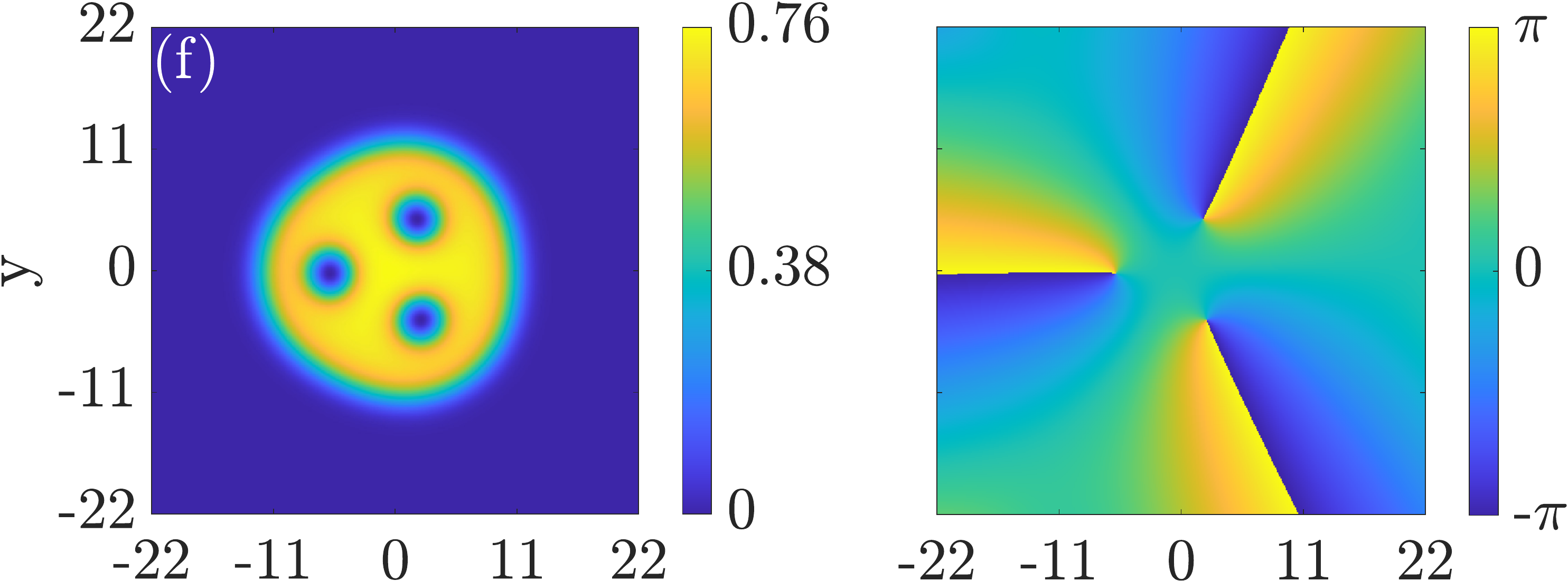}\\
\vspace{0.3\baselineskip}
\includegraphics[width=\columnwidth ,angle=-0]{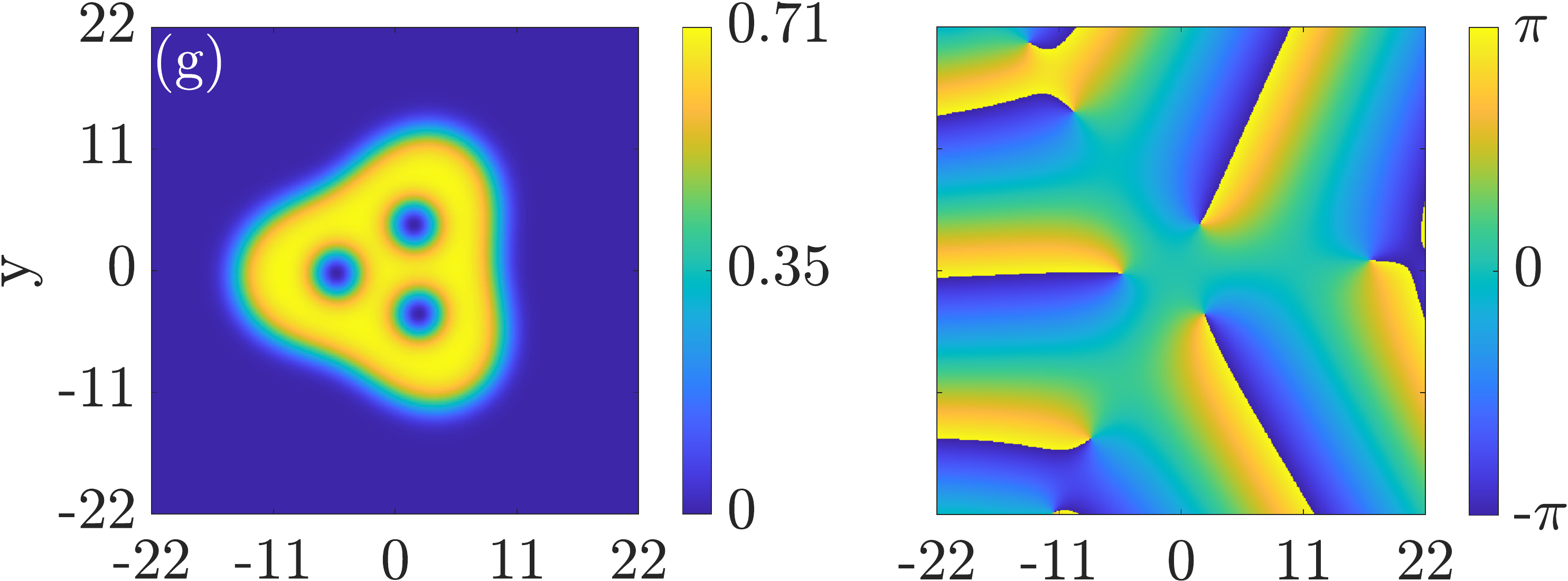}\\
\vspace{0.3\baselineskip}
\includegraphics[width=\columnwidth ,angle=-0]{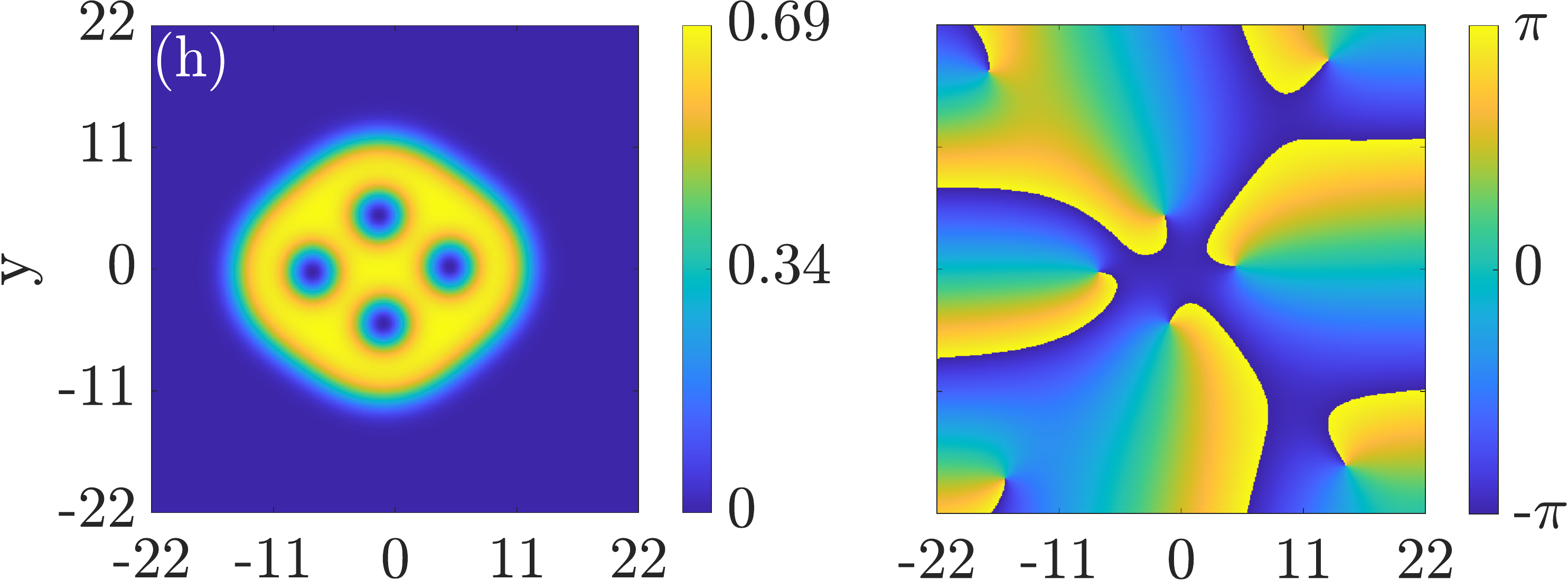}\\
\vspace{0.3\baselineskip}
\includegraphics[width=\columnwidth ,angle=-0]{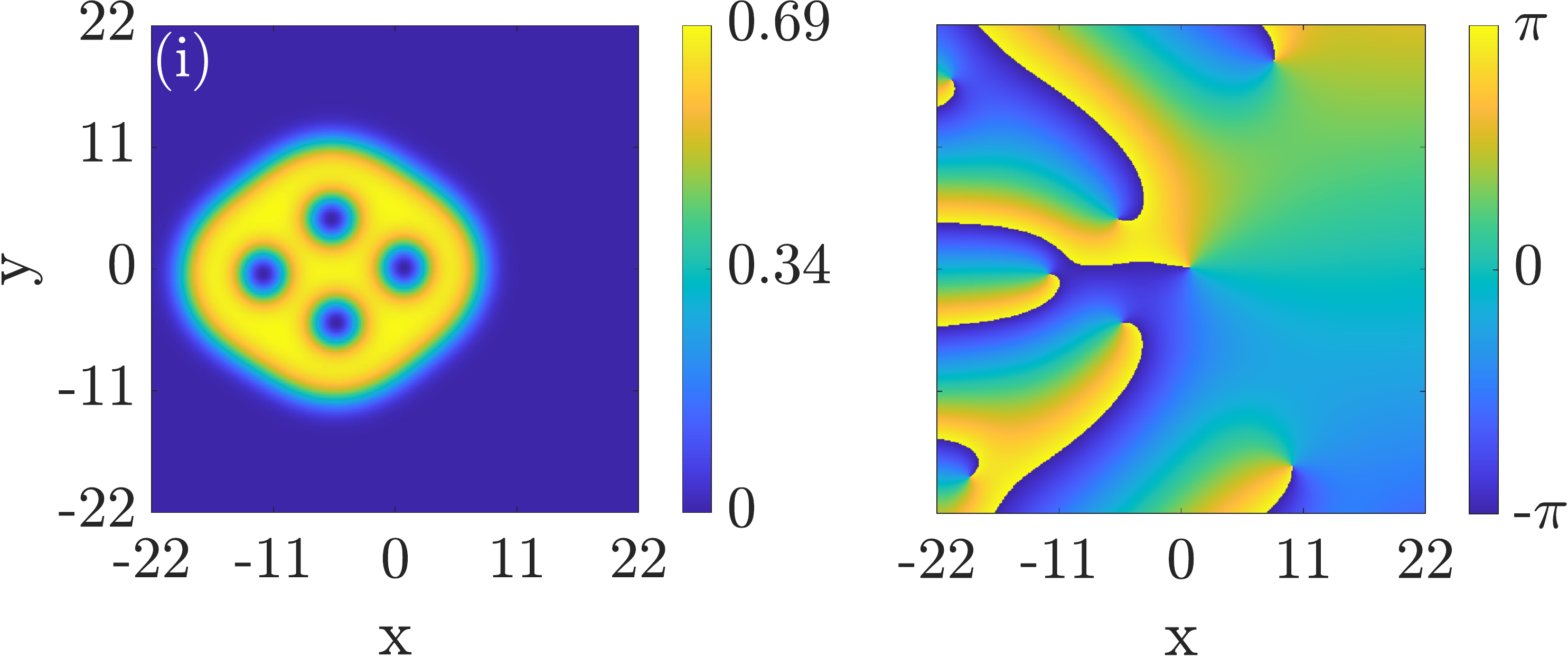}\\
\vspace{\baselineskip}
\includegraphics[width=\columnwidth ,angle=-0]{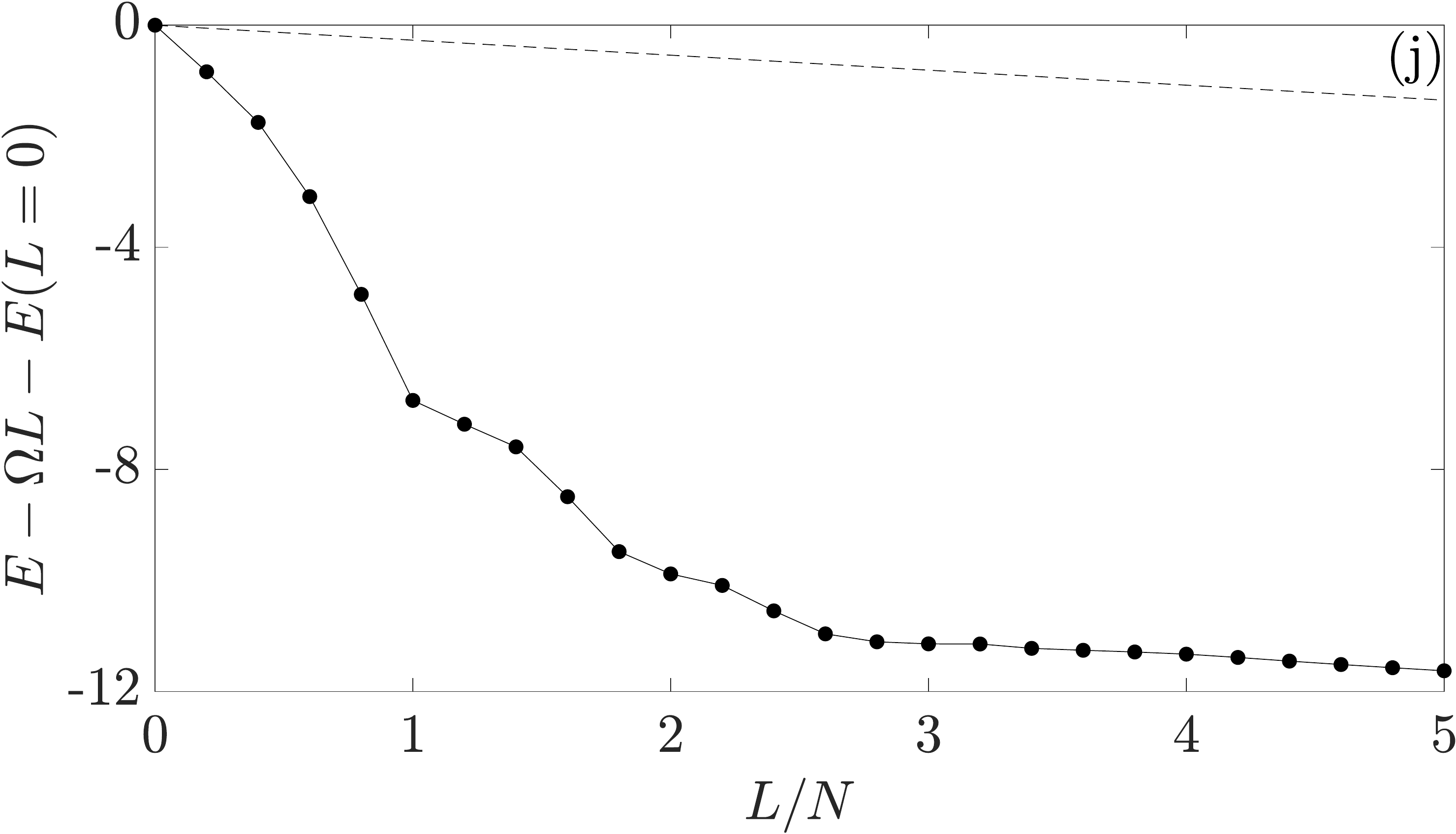}
\caption[]{(Cont.) [(f)\textendash(i)] Same as panels [(a)\textendash(e)] except for $L/N = 2.4, 3.0, 3.4$, and 5.0, respectively. The unit of length is $x_0$. (j) Solid line, with data points: the corresponding dispersion relation in the rotating frame, i.e., $E_{\rm rot}(L/N) - E(L/N = 0)$ as a function of $L/N$, with $\Omega = 0.051$. Dashed line: same as above for the center-of-mass excitation of the nonrotating state. The unit of energy is $E_0$ and the unit of angular momentum is $\hbar$.}
\end{figure}

To get a more quantitative description of the transition from center-of-mass
to vortex excitation, let us consider the eigenfunctions $\phi_m(\rho, \theta)$ 
of the two lowest-Landau levels as trial order parameters for the ground, 
nonrotating state (where $L = 0$), assuming that the oscillator length 
$a_{\rm osc}$ is equal to $\rho_0$,
\begin{eqnarray}
  \phi_{0} = \frac {\sqrt{N}} {\sqrt{\pi} \rho_0} e^{-\rho^2/(2 \rho_0^2)}, 
\end{eqnarray}
and for the state with one singly quantized vortex (where $L = N$),
\begin{eqnarray}
  \phi_{1} = \frac {\sqrt{N}} {\sqrt{\pi} \rho_0^2} \rho e^{i \theta} 
  e^{-\rho^2/(2 \rho_0^2)}.
\end{eqnarray}
Evaluating the energy due to the nonlinear term,
\begin{eqnarray}
  E_{{\rm int},i} = \frac 1 2 \int |\phi_{i}|^4 \ln \frac {|\phi_{i}|^2} 
  {\sqrt e} \, d^2 \rho.
\end{eqnarray} 
For the state $\phi_{0}$ we find
\begin{eqnarray}
  \frac {E_{\rm{int},0}} N = \frac {N} {4 \pi \rho_0^2} 
  \left(\ln \frac N {\pi \sqrt{e} \rho_0^2} - \frac 1 2 \right)
  = \frac {\bar{n}} {4} 
  \left(\ln \frac {\bar{n}} {\sqrt{e}} - \frac 1 2 \right),
\nonumber 
\end{eqnarray} 
while for the state $\phi_{1}$,
\begin{eqnarray}
  \frac {E_{\rm{int},1}} N \approx \frac {N} {2 \pi \rho_0^2} 
  \left(\frac 1 4 \ln \frac N {\pi \sqrt{e} \rho_0^2} - \frac 3 8 + 0.057 \right)
  \nonumber \\
  = \frac {\bar{n}} {2} 
  \left(\frac 1 4 \ln \frac {\bar{n}} {\sqrt{e}} - \frac 3 8 + 0.057 \right).
\end{eqnarray} 
When we have center-of-mass excitation (of the state with $L=0$), from 
Eq.\,(\ref{eqcom}) it follows that 
\begin{eqnarray}
 E_{\rm COM}(L = N) - E(L = 0) = N \omega.
\end{eqnarray}
When we have vortex excitation,
\begin{eqnarray}
 E_{\rm vor}(L = N) - E(L = 0) = N \omega + E_{\rm{int},1} - E_{\rm{int},0}.
\label{evor}
\end{eqnarray}
From the last two equations, we see that it is the difference $E_{\rm{int},1} - 
E_{\rm{int},0}$ which determines whether we will have center-of-mass, or vortex 
excitation. It turns out that the critical value of $N/\rho_0^2$ which gives 
$E_{\rm{int},1} = E_{\rm{int},0}$ is approximately equal to 4. If $\rho_0^2 
= a_{\rm osc}^2 = 1/\omega = 20$, then the critical value of $N$ is approximately 
80. We stress that the calculation presented above compares the energy between 
the ground state and the state with one vortex located at the center of the 
droplet. From our numerical results it follows that, for $\omega = 0.05$, 
the critical number of $N$ for the transition from center-of-mass excitation 
to vortex excitation is between 98.6 and 98.7. 

To examine what happens for even larger values of $N$, we show in 
Fig.\,3 the result of our calculations for $N = 200$ and $\omega = 0.05$, 
i.e., $N \omega = 10$. We observe that for $0 < L < N$ the droplet is again 
distorted from axial symmetry due to the approach of a vortex from infinity [Figs.\,3(b) and 3(c)]. 
When $L = N$ this vortex ends up again at the center of the droplet [Fig.\,3(d)]. However, 
here that the atom number $N$ is larger, for $L > N$ a second vortex enters 
the system, and eventually a twofold symmetric state forms [Figs.\,3(e) to 3(g)]. Here it is only 
for $L/N$ larger than $\approx 2.6$ that the droplet carries its additional 
angular momentum via center-of-mass excitation, i.e., via a ``mixed" state, as shown in Figs.\,3(h) and 3(i). 
The dispersion relation (in the rotating frame), which is also shown in Fig.\,3(j), becomes linear again, now for $L/N$ exceeding $\approx 2.6$.

\subsection{Rotational properties of droplets of ``larger" size}

In Fig.\,4 we have considered an even larger value of $N = 270$, with $\omega$ 
still being equal to 0.05 ($N \omega = 13.5$). Clearly the mean density of the 
nonrotating droplet also increases. As a result, we observe up to four vortices
which are energetically favorable [Figs.\,4(a) to 4(h)], before the ``mixed" state, i.e., the 
center-of-mass excitation of this state with four vortices, becomes the state 
of lowest energy, for $L/N$ exceeding $\approx 3.4$ [Fig.\,4(i)]. As in the case of droplets of ``intermediate" size, the dispersion relation, which is shown in Fig.\,4(j), becomes linear beyond this $L/N$ value.

\subsection{Fixing $\Omega$ instead of $L$}

Up to now all our results have been derived for fixed $L$. From the dispersion 
relation, one may also evaluate the angular momentum of the droplet if $\Omega$ 
is fixed, instead. More specifically, having evaluated the dispersion 
relation (i.e., the lowest energy $E(L)$ as function of $L$), we consider the energy 
in the rotating frame $E_{\rm rot}(L) = E(L) - L \Omega$. For some fixed $\Omega$ we 
find the value of $L$ that minimizes $E_{\rm rot}(L)$ and that is how $L/N(\Omega)$, 
i.e., Fig.\,5, is produced.

\begin{figure}[h]
\includegraphics[width=\columnwidth ,angle=-0]{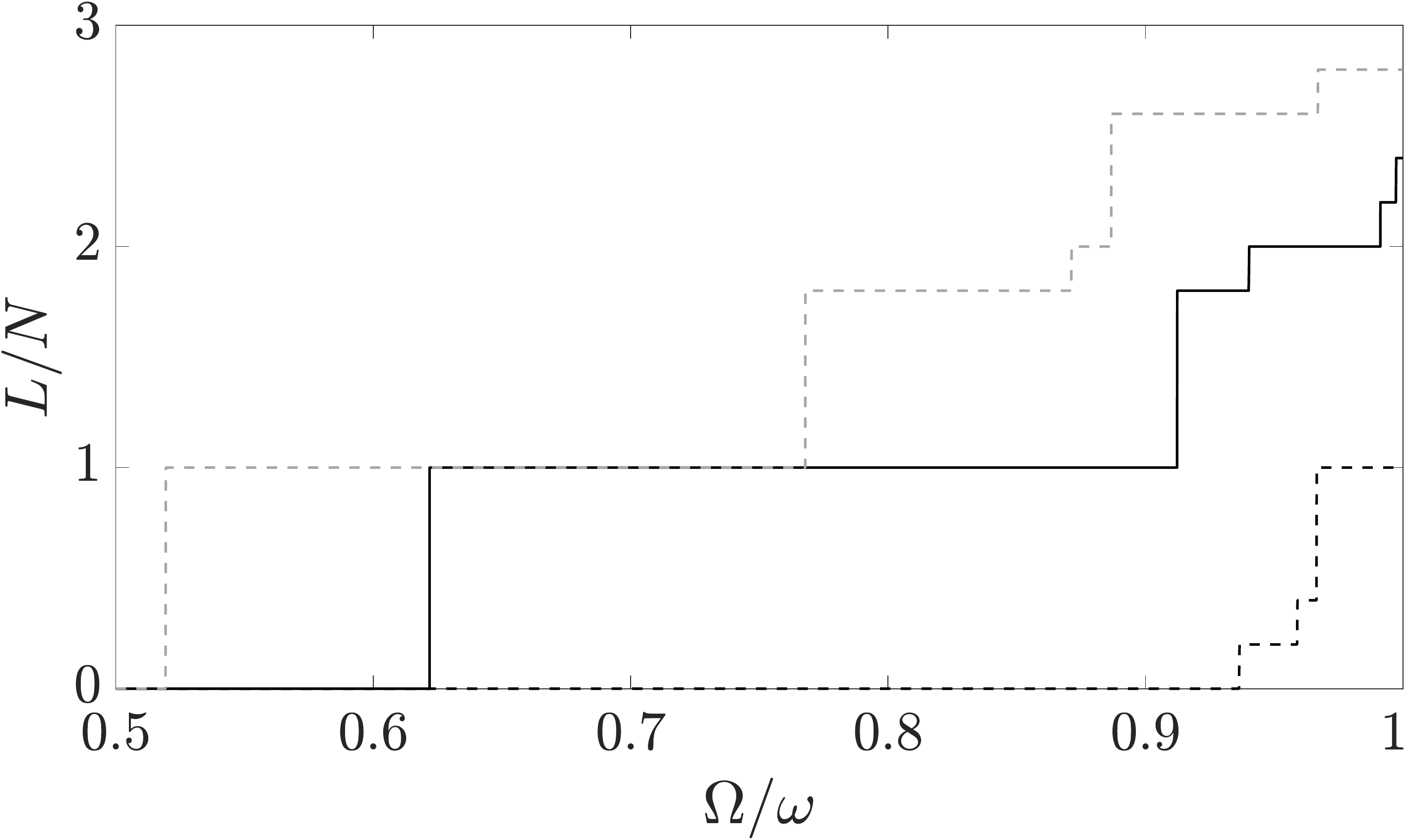}
\caption{The functions $L/N = L/N(\Omega/\omega)$, derived from the lowest-energy states, for $N = 100$ (black, dashed curve), 200 (black, solid curve), and 270 (gray, dashed curve), with $\omega = 0.05$. The unit of angular momentum is $\hbar$.}
\end{figure}

Figure 5 shows $L/N = L/N(\Omega/\omega)$, for $N = 100, 200$, and 270, with $\omega 
= 0.05$ (the steps in the angular momentum per particle $L/N$ that we used to produce 
this plot were equal to 0.2). In this plot we see the usual plateaus, also known in 
the case of single-component condensates with an effectively repulsive contact 
interaction. We stress that for $\Omega \to \omega^-$, this plot diverges, as we argue 
in the following section [see Eq.\,(\ref{dife}) and the relevant discussion].

\section{General picture and limit of rapid rotation}

From the examples presented above, and other cases that we have investigated, 
one may get the more general picture that emerges in this system. For sufficiently 
small $N$ (when $\rho_0 \ll a_{\rm osc}$) we have center-of mass-excitation of the 
nonrotating ground state for all values of $L$. For larger values of $N$, where 
$\rho_0 \gtrsim a_{\rm osc}$, with increasing $L$ one, or more vortices enter the 
cloud. However, there is a limit to this. As the number of vortices increases, 
${\bar n}$ drops. Decreasing $\bar{n}$ even further, is not energetically 
favorable. As a result, if $L$ increases further, the additional angular 
momentum is carried via center-of-mass excitation of some ``mixed" state. 
The dispersion relation also becomes a straight line beyond this specific 
value of $L$.

One estimate for the maximum number of vortices $N_v$ that the droplet 
accommodates before it turns to center-of-mass excitation is that the
mean density is equal to the one that minimizes the energy of 
Eq.\,(\ref{eee}), i.e., ${\bar n} = 1/\sqrt{e}$,
\begin{eqnarray}
 \frac N {S - N_v \sigma} = \frac 1 {\sqrt e}.
\end{eqnarray}
Here $S$ and $\sigma$ are the ``surfaces" of the droplet and of each vortex, 
respectively. An approximate expression for $\sigma$ is $\sigma \approx \pi 
\xi^2$, where the coherence length $\xi$ gives roughly the linear size of the 
vortex.

According to the analysis presented above, one may also make a general 
statement about the dispersion relation. For any two states with angular 
momentum $L_1$ and $L_2$, with $L_1 < L_2$, $E(L_2)$ has to be lower than 
$E(L_1) + (L_2 - L_1) \omega$,
\begin{eqnarray}
 E(L_2) < E(L_1) + (L_2 - L_1) \omega.
\label{erot}
\end{eqnarray} 
If this inequality is violated, one may always start with the state of 
angular momentum $L_1$ and excite it via center-of-mass excitation to a
state with angular momentum $L_2$. In this case, $E(L_2)$ will be equal 
to $E(L_1) + (L_2 - L_1) \omega$. From Eq.\,(\ref{erot}) it also follows 
that, for $L_2 \to L_1$,
\begin{eqnarray}
 \frac {d E(L)} {d L} < \omega,
\end{eqnarray}
i.e., the slope of the dispersion relation cannot exceed $\omega$.

Another consequence of Eq.\,(\ref{erot}) is that, if one works with a fixed 
rotational frequency of the trap $\Omega$ and not with a fixed angular 
momentum, $\Omega$ cannot exceed $\omega$. Indeed, according to 
Eq.\,(\ref{erot}),
\begin{eqnarray}
  E_{\rm rot}(L_2) < E_{\rm rot}(L_1) + (L_2 - L_1) (\omega - \Omega).
\label{dife}
\end{eqnarray}
Therefore, if $\Omega \ge \omega$, $E_{\rm rot}(L_2) < E_{\rm rot}(L_1)$ 
and $E_{\rm rot}(L)$ is a decreasing function of $L$. In other words, if 
$\Omega$ exceeds $\omega$, then the energy is unbounded. This result is 
a combined effect of the ``mixed" state that we have seen, with the 
centrifugal force, which gives rise to the effective potential 
$M (\omega^2 - \Omega^2) \rho^2/2$. Last but not least, we stress that 
this result is also true in the case of contact interactions, in a harmonic 
trapping potential. 

\begin{figure}
\includegraphics[width=\columnwidth ,angle=-0]{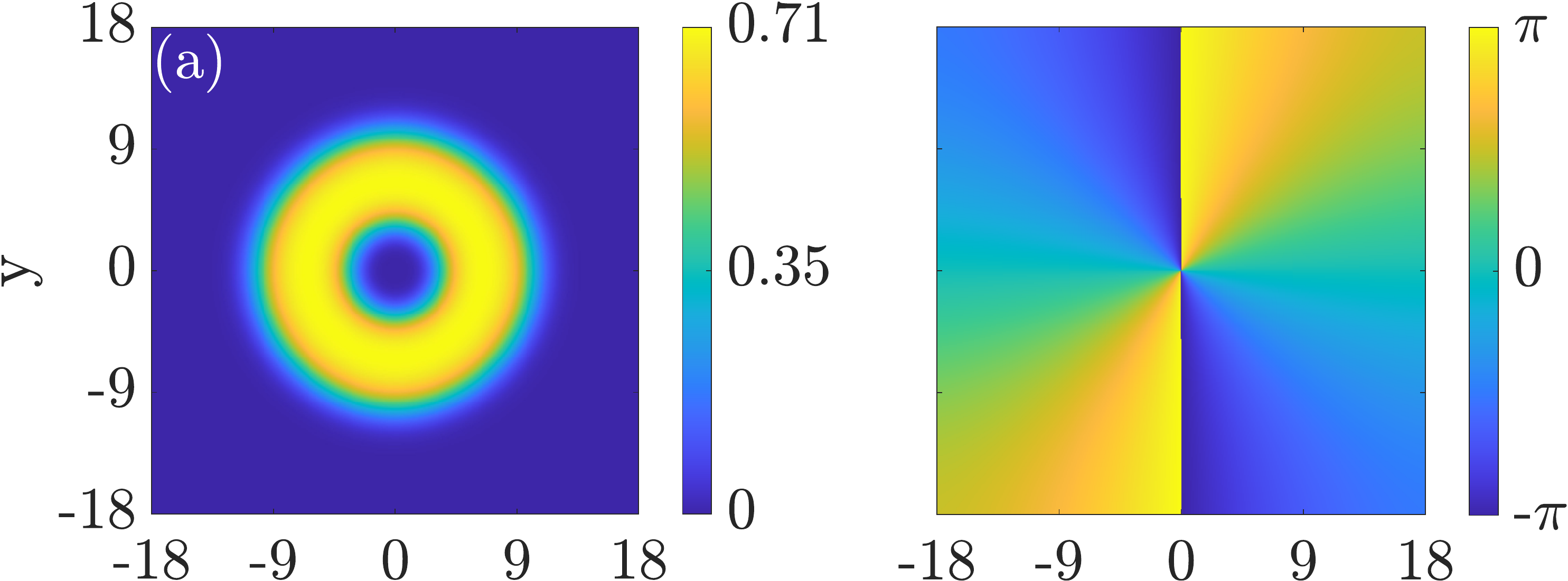}\\
\vspace{0.3\baselineskip}
\includegraphics[width=\columnwidth ,angle=-0]{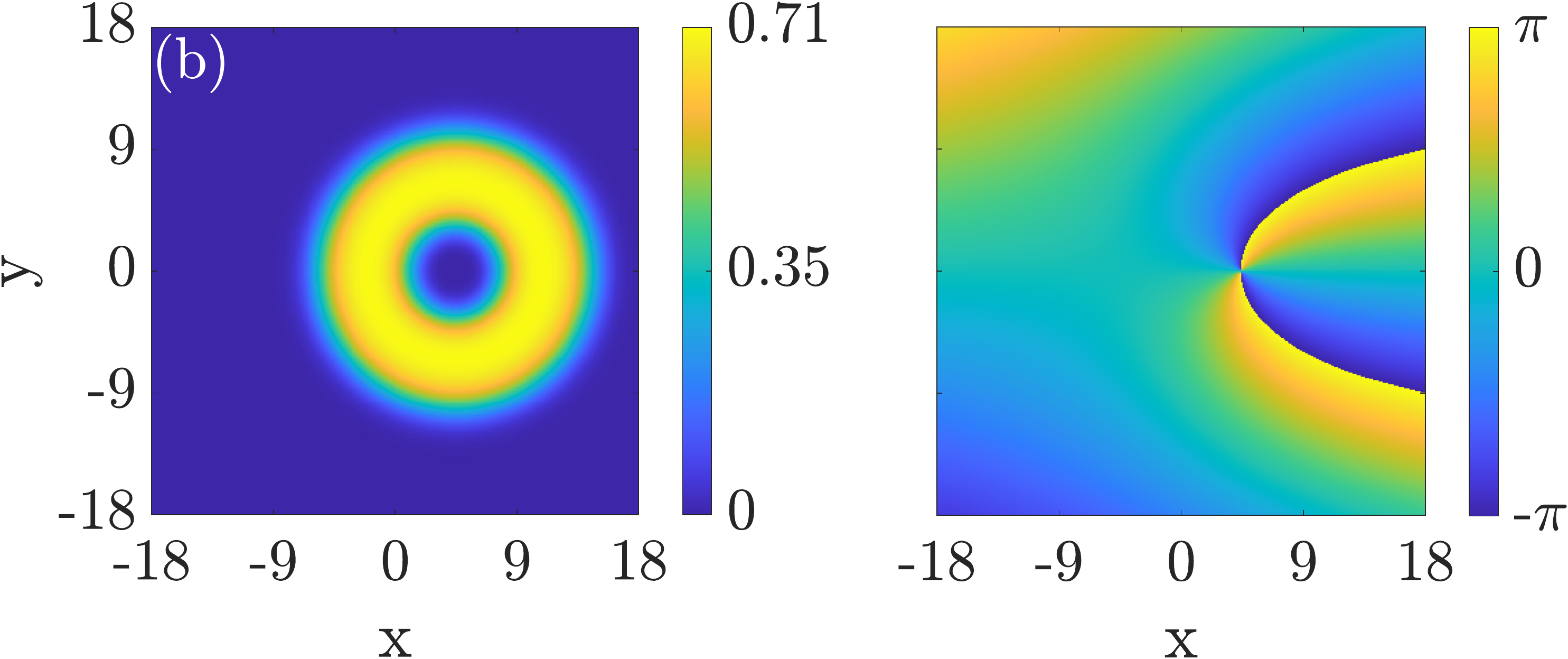}\\
\vspace{\baselineskip}
\includegraphics[width=\columnwidth ,angle=-0]{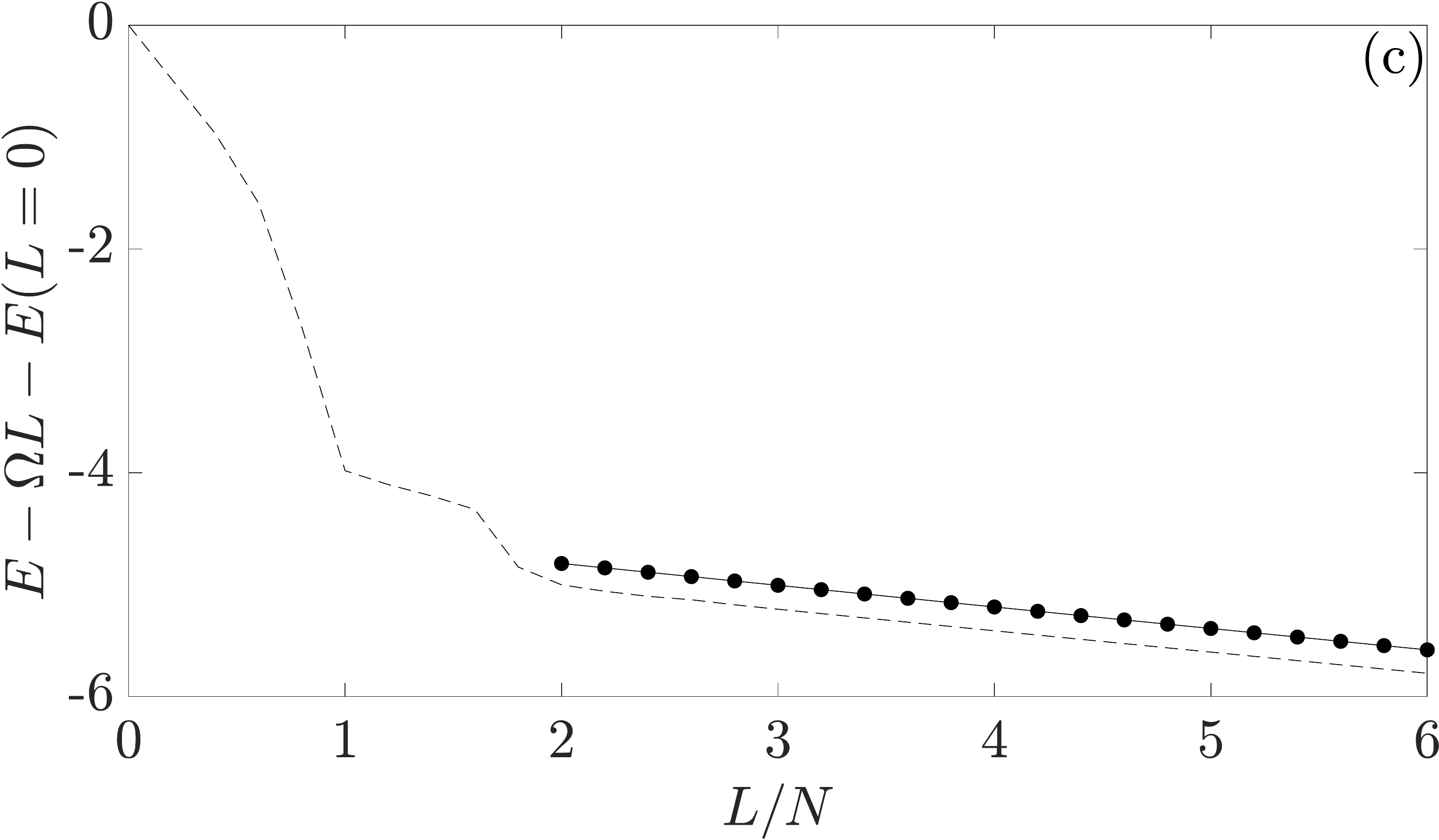}
\caption{[(a), (b)] The density (left column, in units of $\Psi_0^2$) and the phase (right column) of the droplet order parameter, in the excited, multiply quantized vortex state(s) $\Psi_{S=2}$, for $N = 200$, $\omega = 0.05$, and $L/N = 2.0$ and 3.0, respectively. The unit of length is $x_0$. (c) Solid line, with data points: the corresponding dispersion relation in the rotating frame, i.e., $E_{\rm rot}(L/N) - E(L/N = 0)$ as a function of $L/N$, with $\Omega = 0.051$. Dashed line: same as above for the lowest-energy state. The unit of energy is $E_0$ and the unit of angular momentum is $\hbar$.}
\end{figure}

\section{Excitation spectrum}

\begin{figure}[!ht]
\includegraphics[width=\columnwidth ,angle=-0]{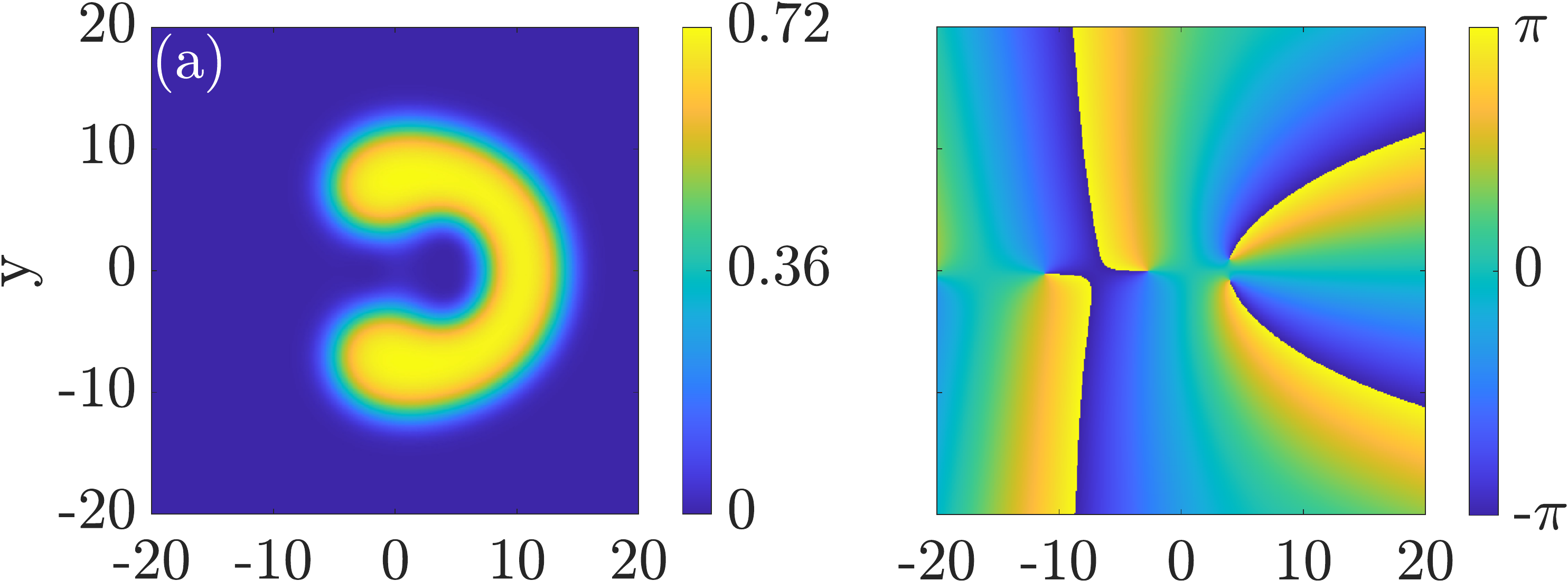}\\
\vspace{0.3\baselineskip}
\includegraphics[width=\columnwidth ,angle=-0]{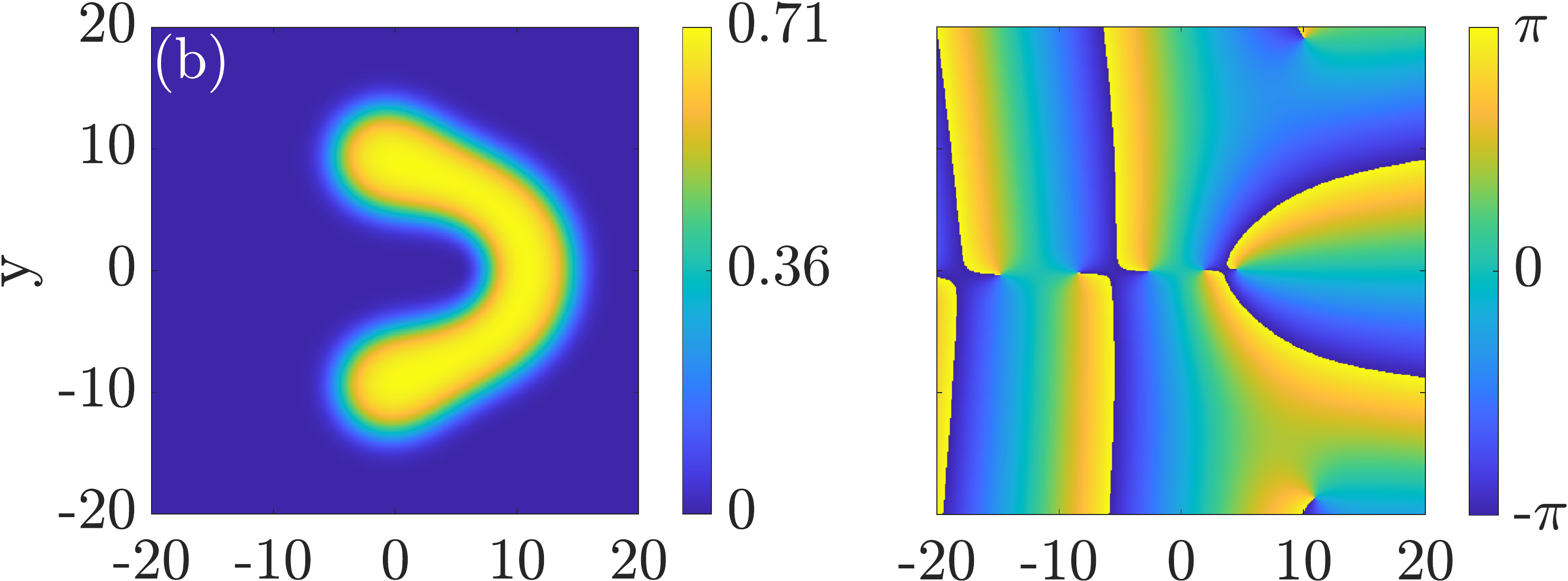}\\
\vspace{0.3\baselineskip}
\includegraphics[width=\columnwidth ,angle=-0]{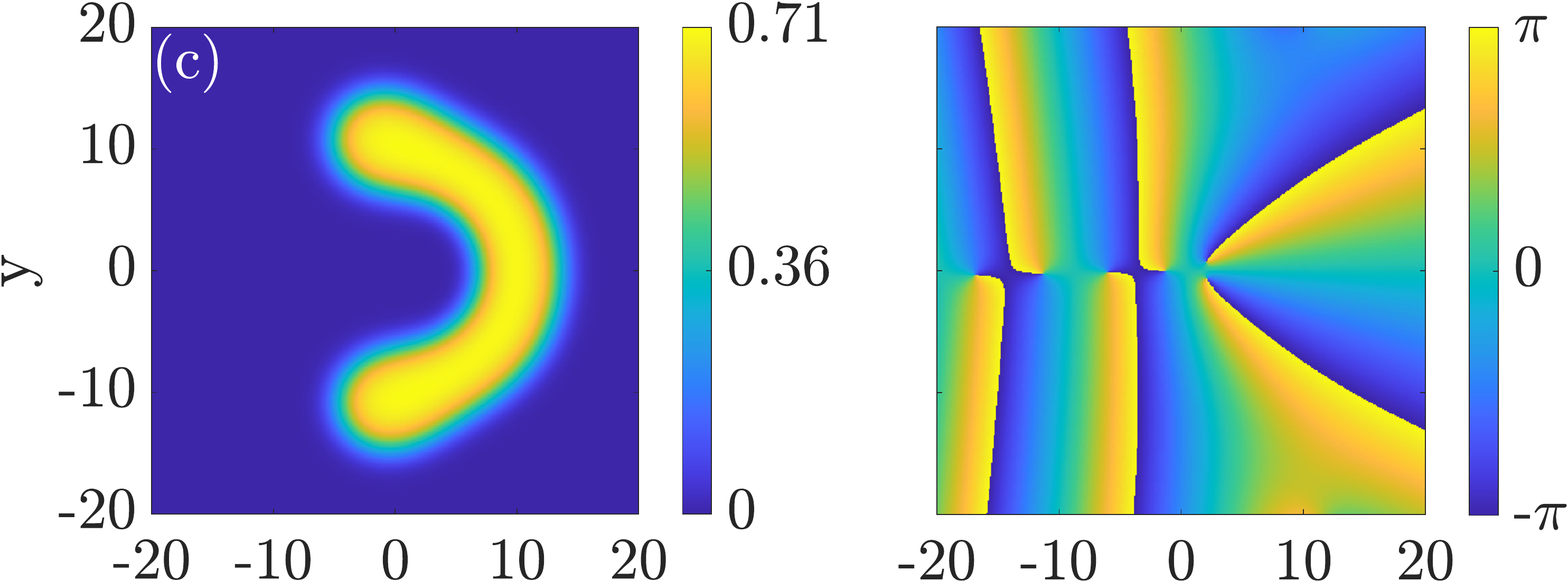}\\
\vspace{0.3\baselineskip}
\includegraphics[width=\columnwidth ,angle=-0]{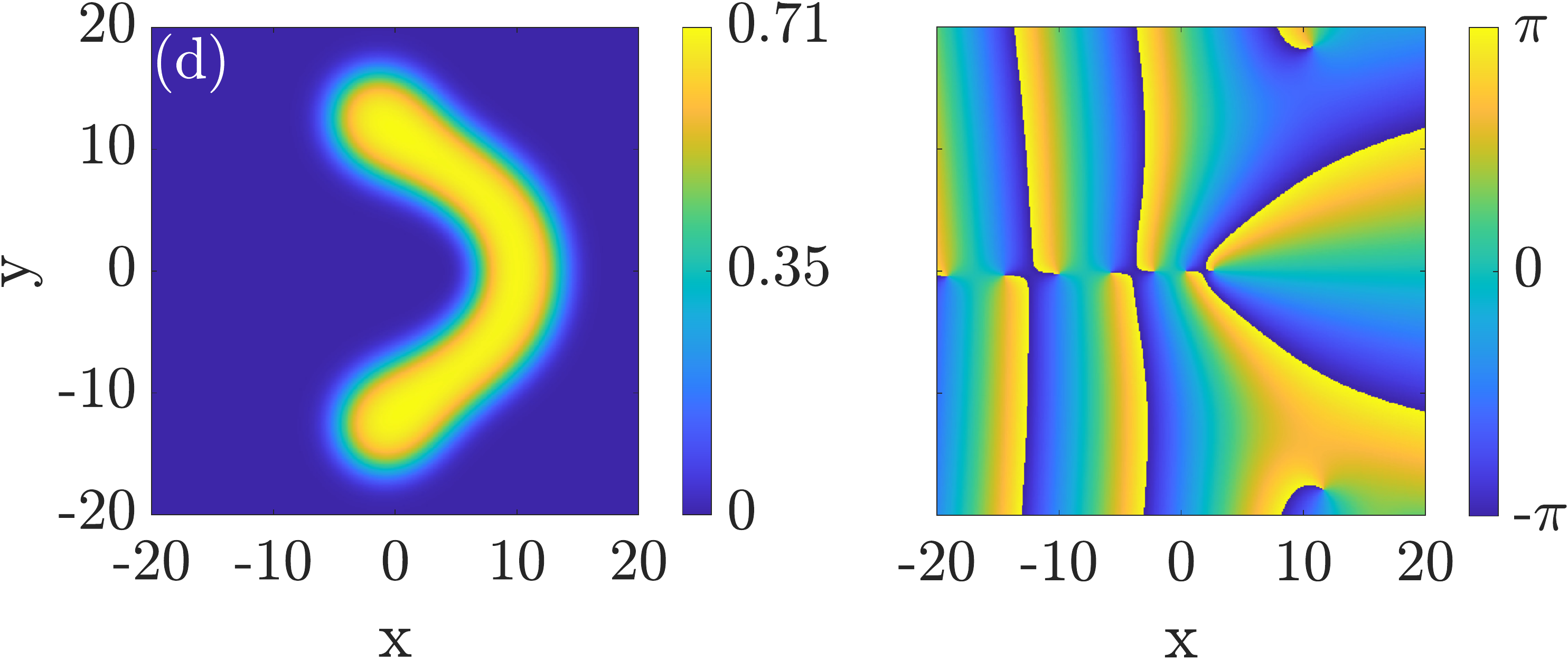}\\
\vspace{\baselineskip}
\includegraphics[width=\columnwidth ,angle=-0]{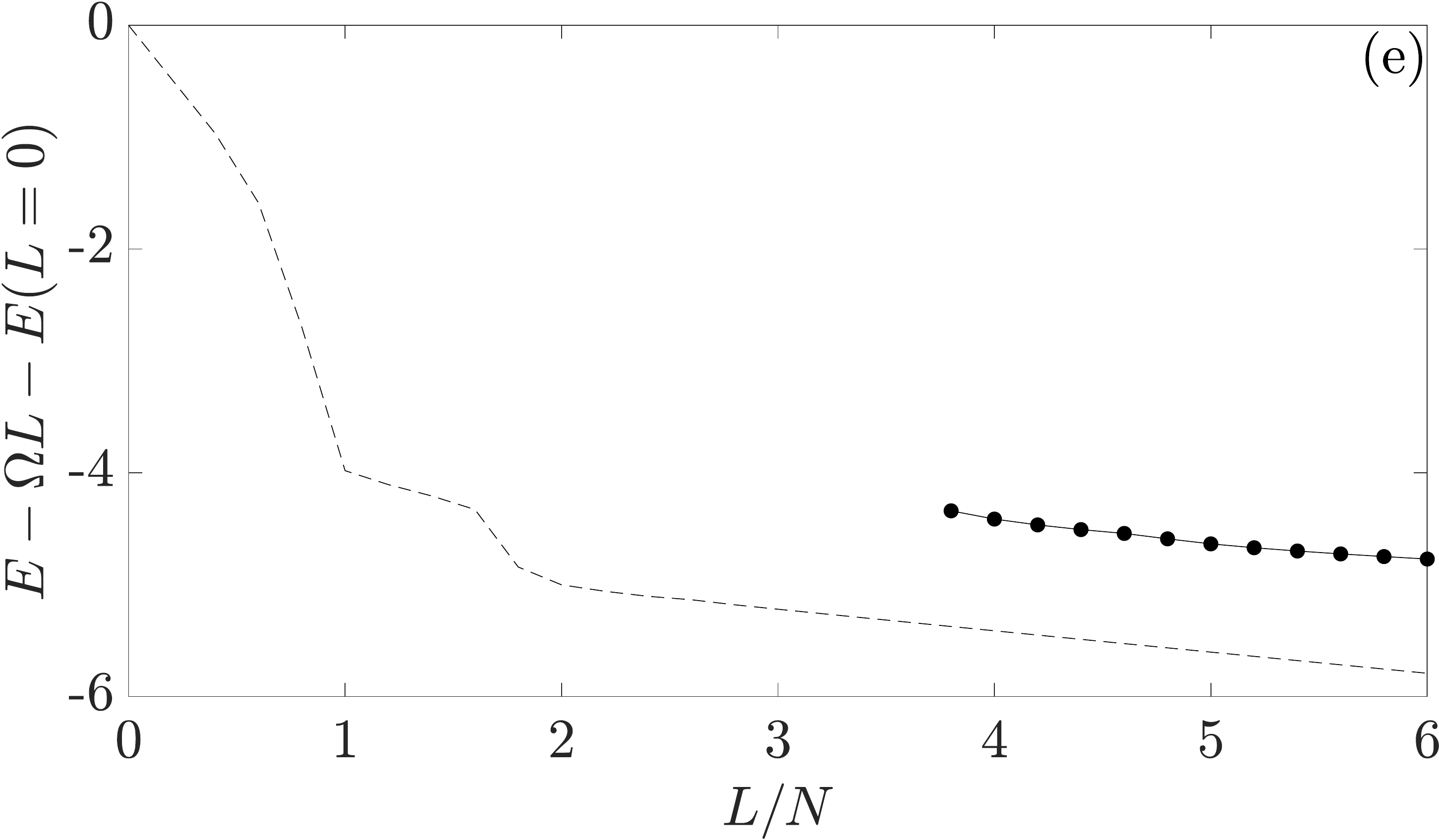}
\caption{[(a)\textendash(d)] The density (left column, in units of $\Psi_0^2$) and the phase (right column) of the droplet order parameter, in the excited states with an axially asymmetric density distribution for $N = 200$, $\omega = 0.05$, and $L/N = 3.8, 4.4, 5.0$, and 5.6, respectively. The unit of length is $x_0$. (e) Solid line, with data points: the corresponding dispersion relation in the rotating frame, i.e., $E_{\rm rot}(L/N) - E(L/N = 0)$ as function of $L/N$, with $\Omega = 0.051$. Dashed line: same as above for the lowest-energy state. The unit of energy is $E_0$ and the unit of angular momentum is $\hbar$.}
\end{figure}

All the states that we have presented so far are those of lowest energy, 
for a fixed $N$ and $L$. Although this is one of the most important questions, 
a separate question is the excitation spectrum. We should stress that the 
excitation spectrum is not only interesting theoretically, but is also 
experimentally relevant. While we have not made a complete study of the 
excited states, we have managed to find at least part of them. Interestingly 
enough, the arguments presented in Sec.\,IV allow us to get a rather easy 
understanding of this problem and to even predict the existence of the states
that we have identified.

In the results which are presented below, we have focused on the case $N = 200$ 
and $\omega = 0.05$ and we have identified two classes of states in the excitation 
spectrum. The first class includes multiply quantized vortex states, of the form 
$\Psi_S(\rho, \theta) = f(\rho) e^{i S \theta}$, where $S$ is the winding number,
which have an axially symmetric density distribution. These are solutions of the 
equation 
\begin{eqnarray}
 - \frac 1 2 \frac {\partial^2 f} {\partial \rho^2} 
 - \frac 1 {2 \rho} \frac {\partial f} {\partial \rho}
 + \frac {S^2} {2 \rho^2} f + \frac 1 2 \omega^2 \rho^2 f +
\nonumber \\
 + |f|^2 \ln |f|^2 f = \mu f.
\label{mvor}
\end{eqnarray}
Starting with $L/N = S = 2$, we have found that this doubly quantized vortex 
state [Fig.\,6(a)], $\Psi_{S=2}$, is very close in energy with the actual state of lowest 
energy, as shown in Fig.\,6(c). This proximity is not a surprise, but rather is 
expected, i.e., it is due to the fact that the mean densities of the two states 
are very close to each other. For $L/N > 2$, we then have center-of-mass 
excitation of the doubly quantized vortex state [Fig.\,6(b)], with an energy which increases 
linearly with the angular momentum, as we saw earlier. Clearly what we described 
for $L/N \ge 2$ is general. For example, the state $\Psi_{S=3}$ is also present
in the excitation spectrum for all values of $L/N \ge 3$, etc. 

The multiply quantized vortex states described above have an axially symmetric 
density distribution with respect to their center of mass. The second class of 
states that we have identified in the excitation spectrum, are states which break 
the axial symmetry of the problem. In this case the centrifugal term [i.e., the 
third term on the left in Eq.\,(\ref{mvor})] favors an axially asymmetric density 
distribution. As a result, the cloud ``localizes", since this is energetically more 
favorable (in order, again, for the droplet to achieve the optimal mean density). 
Examples of such excited states are shown in Figs.\,7(a) to 7(d), as well as the corresponding energy in Fig.\,7(e).

\section{Physical units and experimental relevance of our results}

As mentioned above, up to now we have used dimensionless units. Here we show
how one may return to the physical units and then we give some estimates for 
the experimentally relevant scales.

First of all, let us denote as $\Psi_{\uparrow}$ and $\Psi_{\downarrow}$ the order parameter
of each component. In the symmetric case that we consider in the present problem, $\Psi_{\uparrow}
= \Psi_{\downarrow}$ and also $\int |\Psi_{\uparrow}|^2 d^2 \rho = \int |\Psi_{\uparrow}|^2 d^2 \rho 
= N/2$, where $N$ is the total number of atoms in both components. Let us also introduce
$\Psi = {\sqrt 2} \Psi_{\uparrow} = {\sqrt 2} \Psi_{\downarrow}$, where obviously
$\int |\Psi|^2 d^2 \rho = N$. 

The order parameter $\Psi$ satisfies the equation
\begin{eqnarray}
  i \hbar \frac {\partial \Psi} {\partial t} = - \frac {\hbar^2} {2 M} \nabla ^2 \Psi  
  +  \frac 1 2 M \omega^2 \rho^2 \Psi + 
  \nonumber \\
  + \frac {4 \pi \hbar^2} {M \ln^2(a_{\uparrow \downarrow}/a)} |\Psi|^2 
  \ln \frac {|\Psi|^2} {2 {\sqrt e} n_0} \Psi.
  \label{eq1}
\end{eqnarray}
Here $M$ is the atom mass, which is assumed to be the same for the two components and 
$\omega$ is the frequency of the (two-dimensional) trapping potential. Also, $a$ and 
$a_{\uparrow \downarrow}$ are the two-dimensional scattering lengths for elastic 
atom-atom collisions between the same species (assumed to be equal for the two 
components) and for different species, respectively. Furthermore,
\begin{equation}
   n_0 = \frac {e^{-2 \gamma - 3/2}} {2 \pi} \frac {\ln(a_{\uparrow \downarrow}/a)} 
   {a a_{\uparrow \downarrow}}.
\end{equation}
Here $\gamma$ is Euler's constant, $\gamma \approx 0.5772$, while 
\begin{eqnarray}
    \ln (a_{\uparrow \downarrow}/a) = \sqrt{\frac {\pi} 2} 
    \left( \frac {a_z} {a^{\rm 3D}} - \frac {a_z} {a_{\uparrow \downarrow}^{\rm 3D}} \right).
\end{eqnarray}
Here $a_z$ is the ``width" of the droplet along the axis of rotation, and $a^{\rm 3D}$, 
$a_{\uparrow \downarrow}^{\rm 3D}$ are the three-dimensional scattering lengths for  
elastic atom-atom collisions between the same and different species, respectively. 
Introducing 
\begin{eqnarray}
  \Psi_0^2 = 2 \sqrt{e} n_0 = \frac {e^{-2 \gamma - 1}} {\pi} \frac {\ln(a_{\uparrow \downarrow}/a)} 
   {a a_{\uparrow \downarrow}}
\end{eqnarray}
and setting ${\tilde \Psi} = \Psi/\Psi_0$, Eq.\,(\ref{eq1}) becomes
\begin{eqnarray}
  i \frac {\partial {\tilde \Psi}} {\partial \tilde{t}} = 
  - \frac {1} {2} {\tilde \nabla}^2 {\tilde \Psi}  
  +  \frac 1 2 {\tilde \omega}^2 {\tilde \rho}^2 {\tilde \Psi} 
  + |{\tilde \Psi}|^2 \ln |{\tilde \Psi}|^2 {\tilde \Psi}.
  \label{eq2}
\end{eqnarray}
Here $\tilde{t} = t/t_0$, where 
\begin{equation}
 t_0 = \frac {M a a_{\uparrow \downarrow} \ln(a_{\uparrow \downarrow}/a)} {4 \hbar e^{-2 \gamma - 1}}. 
\end{equation}
Also, ${\tilde \rho} = \rho/x_0$ and ${\tilde \nabla}^2$ is the dimensionless Laplacian, with the 
unit of length being $x_0$, where
\begin{equation}
 x_0 = \sqrt{\frac {a a_{\uparrow \downarrow} \ln(a_{\uparrow \downarrow}/a)} 
 {4 e^{-2 \gamma - 1}}}.
\end{equation}
Furthermore, ${\tilde \omega} = \omega/\omega_0$, where the units of the frequency $\omega_0$ 
and of the energy $E_0$, are
\begin{equation}
  E_0 = \hbar \omega_0 = \frac {\hbar} {t_0} 
  = \frac {\hbar^2} {M x_0^2} = \frac {\hbar^2} {M a a_{\uparrow \downarrow}} 
  \frac {4 e^{-2 \gamma - 1}} {\ln(a_{\uparrow \downarrow}/a)}.
\end{equation}
The normalization condition takes the form
\begin{eqnarray}
 \int |{\tilde \Psi}|^2 \, d^2 {\tilde \rho} = \frac N {N_0},
\end{eqnarray}
where 
\begin{equation}
  N_0 = \Psi_0^2 x_0^2 = \frac 1 {4 \pi} \ln^2(a_{\uparrow \downarrow}/a),
\label{n0}
\end{equation}
which is the unit of $N$. 

Finally, the time-independent equation that corresponds to Eq.\,(\ref{eq2}) is derived after 
we set $\Psi({\tilde{\bf \rho}}, {\tilde t}) = \Psi({\tilde{\bf \rho}}) e^{- i {\tilde \mu} 
\tilde{t}}$, where ${\tilde \mu}$ is the dimensionless chemical potential, thus getting
\begin{eqnarray}
  - \frac {1} {2} {\tilde \nabla}^2 {\tilde \Psi} 
  + \frac 1 2 {\tilde \omega}^2 {\tilde \rho}^2 {\tilde \Psi} 
  +  |{\tilde \Psi}|^2 \ln |{\tilde \Psi}|^2 {\tilde \Psi} = {\tilde{\mu}} {\tilde \Psi}.
  \label{eq22}
\end{eqnarray}
We stress that the ``tilde" used in the symbols in the present section, which
represents dimensionless quantities, is dropped in all the other sections for convenience. 

Equation (\ref{n0}) allows us to evaluate the actual (total) number of atoms in a droplet. 
For a typical value of $a_z = 0.1$ $\mu$m and $a^{\rm 3D} = 10.1$ nm, $a_{\uparrow 
\downarrow}^{\rm 3D} = -10.0$ nm, $\ln (a_{\uparrow \downarrow}/a) \approx 25$. Then, 
according to Eq.\,(\ref{n0}), $N_0 \approx 50$. Therefore, the range of $N$ that we 
have considered (50 up to 270) corresponds roughly to $\approx 2500$, up to $\approx 
14000$ atoms in an experiment.

Also, the unit of length $x_0$ turns out to be on the order of 1 $\mu$m.
This implies that, for e.g., $10^4$ atoms, the size of a (nonrotating) droplet 
in the Thomas-Fermi limit, which was evaluated in Sec.\,III, is $\approx 10$ $\mu$m.
Finally, typical values of the two-dimensional density are $\approx 10^9$ ${\rm cm}^{-2}$, 
of the three-dimensional density are $10^{13}$ ${\rm cm}^{-3}$, $t_0$ is on the order of 
millisecond and the typical value of the trapping potential is hundreds of hertz.

\section{Summary of the results with a comparison with the problem of contact interactions}

In the present study we investigated the rotational behavior of a quasi-two-dimensional 
quantum droplet, which consists of a mixture of two distinguishable 
Bose-Einstein condensed gases, assuming that the droplet is confined in a harmonic 
trapping potential. 

For a fixed trap frequency and sufficiently small atom numbers, the droplet
does not host any vortices, but rather it carries its angular momentum via 
center-of-mass excitation of its nonrotating, ground state. This is very much 
like the case of a single-component Bose-Einstein condensed gas, which has an 
effectively attractive interatomic interaction potential and is confined in a 
harmonic trap. The only difference between the two problems is that, while 
in the case of droplets we have a stable system (as a consequence of quantum 
fluctuations), in the case of a single component the system is metastable.

For a larger atom number, and sufficiently small values of the angular momentum,
the droplet behaves in the usual way, with vortices entering it as the angular
momentum increases. As more and more vortices enter the droplet, its average 
density drops, which is energetically favorable. However, as the number of
vortices increases, eventually it is no longer energetically favorable for 
even more vortices to enter the droplet. As a result, beyond some critical 
value of the angular momentum the droplet carries the additional angular 
momentum via center-of-mass excitation of a vortex-carrying state. 

For a single-component, harmonically trapped Bose-Einstein 
condensate with an effectively attractive interaction the angular momentum 
is carried via center-of-mass excitation of the nonrotating state, for all 
values of the angular momentum. On the contrary, for an effectively repulsive 
interaction this never happens (in the lowest-energy state) \cite{finiten}. 
Furthermore, for a contact potential with an effective repulsive interaction, 
the interaction energy is a decreasing function of the density.

In the case of a two-component system, i.e., in quantum droplets, the situation 
is different due to a simple and important difference between the two problems. 
Here, the interaction energy is not a monotonic function of the density [see 
Eq.\,(\ref{eee})], but rather it has a minimum at some specific value of the 
density.

As a result, as $L$ increases, in the case of a contact potential with an 
effective repulsive interaction, the cloud expands radially and this lowers 
its mean density and the corresponding interaction energy. Eventually, the 
system enters the highly correlated ``Laughlin-like" regime that we mentioned 
in the Introduction. On the other hand, for the case of droplets (i.e., 
two-component systems), the decrease of the mean density due to the vortices 
\textemdash for a sufficiently large atom number \textemdash is energetically favorable only 
until the density reaches some finite value.

The important conclusion that follows from the above discussion is the following: 
For increasing $L$, in a single-component condensate the gas enters the 
highly correlated Laughlin regime. On the other hand, when we have two components,
i.e., in the case of droplets, for a sufficiently large angular momentum, a droplet 
is always in a ``mixed" state, i.e., in a state of center-of-mass excitation 
of a state which includes vortices.

Our study demonstrates the richness of this problem, in terms of the various 
physical states. In addition, it also demonstrates that, despite the difference 
of the phases that we have found, there is a universal behavior of the droplets 
in the limit of rapid rotation, in a ``mixed" state, which has never been 
seen before in any other ``traditional" superfluid, including liquid helium and 
harmonically trapped condensed atoms interacting with contact interactions.


\begin{thebibliography}{55}
\bibitem{finiten} In the case of effectively repulsive contact interactions, 
only in one case the state with center-of-mass excitation appears as the state
of lowest energy. This is for a finite number of atoms, where the many-body 
state with $L = 1$ results from center-of-mass excitation of the nonrotating 
many-body state, with $L = 0$ \cite{BM}.
\end{thebibliography}
\end{document}